\documentclass[onecolumn]{IEEEtran}
\usepackage{multicol}
\usepackage{times,amssymb,amsmath,amsfonts,color,cite,bbm,mathrsfs,float,stmaryrd}
\usepackage{graphicx}
\usepackage{enumerate,bm}
\usepackage{latexsym}
\usepackage{theorem}
\usepackage[linesnumbered,ruled]{algorithm2e}
\usepackage[normalem]{ulem}
\usepackage{hyperref}
\usepackage{cleveref}

\newtheorem{theorem}{Theorem}[section]
\newtheorem{lemma}[theorem]{Lemma}
\newtheorem{claim}[theorem]{Claim}

\newtheorem{proposition}[theorem]{Proposition}

\newtheorem{corollary}[theorem]{Corollary}
\newtheorem{definition}[theorem]{Definition}

\newtheorem{observation}[theorem]{Observation}

\newtheorem{example}[theorem]{Example}
\newtheorem{remark}[theorem]{Remark}

\newcommand{\Inv}[1]{\textup{Inv}\left(#1\right)}
\newcommand{\Syn}[1]{\textup{Syn}\left(#1\right)}
\newcommand{\Rnum}[1]{\lowercase\expandafter{\romannumeral #1\relax}}
\newcommand{\RNum}[1]{\uppercase\expandafter{\romannumeral #1\relax}}
\newcommand{\tabincell}[2]{\begin{tabular}{@{}#1@{}}#2\end{tabular}}
\newcommand{\Red}[2]{\textup{Red}_{#1}\left(#2\right)}
\newcommand{\red}[1]{\textup{Red}_{#1}}
\newcommand{\Expan}[2]{\textup{Expan}_{#1}\left(#2\right)}
\newcommand{\Enc}[1]{\textup{Enc}\left(#1\right)}

\renewcommand{\thefigure}{\thesection.\arabic{figure}}
\usepackage{amsmath}

\allowdisplaybreaks

\makeatletter
\renewcommand{\@endtheorem}{\endtrivlist}
\makeatother

\newcommand\remove[1]{}

\makeatletter
\renewcommand{\thefigure}{{\@arabic\c@figure}}
\renewcommand{\fnum@figure}{{\bf Figure\,\thefigure}}
\makeatother

\newcommand\nc\newcommand
\nc{\cA}{\mathcal{A}}\nc{\cB}{\mathcal{B}}\nc{\cC}{\mathcal{C}}\nc{\cD}{\mathcal{D}}
\nc{\cE}{\mathcal{E}}\nc{\cF}{\mathcal{F}}\nc{\cG}{\mathcal{G}}\nc{\cH}{\mathcal{H}}
\nc{\cI}{\mathcal{I}}\nc{\cJ}{\mathcal{J}}\nc{\cK}{\mathcal{K}}\nc{\cL}{\mathcal{L}}
\nc{\cM}{\mathcal{M}}\nc{\cN}{\mathcal{N}}\nc{\cO}{\mathcal{O}}\nc{\cP}{\mathcal{P}}
\nc{\cQ}{\mathcal{Q}}\nc{\cR}{\mathcal{R}}\nc{\cS}{\mathcal{S}}\nc{\cT}{\mathcal{T}}
\nc{\cU}{\mathcal{U}}\nc{\cV}{\mathcal{V}}\nc{\cW}{\mathcal{W}}\nc{\cX}{\mathcal{X}}
\nc{\cY}{\mathcal{Y}}\nc{\cZ}{\mathcal{Z}}

\nc{\bba}{\mathbf{a}}\nc{\bbb}{\mathbf{b}}\nc{\bbc}{\mathbf{c}}\nc{\bbd}{\mathbf{d}}
\nc{\bbe}{\mathbf{e}}\nc{\bbf}{\mathbf{f}}\nc{\bbg}{\mathbf{g}}\nc{\bbh}{\mathbf{h}}
\nc{\bbi}{\mathbf{i}}\nc{\bbj}{\mathbf{j}}\nc{\bbk}{\mathbf{k}}\nc{\bbl}{\mathbf{l}}
\nc{\bbm}{\mathbf{m}}\nc{\bbn}{\mathbf{n}}\nc{\bbo}{\mathbf{o}}\nc{\bbp}{\mathbf{p}}
\nc{\bbq}{\mathbf{q}}\nc{\bbr}{\mathbf{r}}\nc{\bbs}{\mathbf{s}}\nc{\bbt}{\mathbf{t}}
\nc{\bbu}{\mathbf{u}}\nc{\bbv}{\mathbf{v}}\nc{\bbw}{\mathbf{w}}\nc{\bfx}{\mathbf{x}}
\nc{\bby}{\mathbf{y}}\nc{\bbz}{\mathbf{z}}

\nc{\bbA}{\mathbf{A}}\nc{\bbB}{\mathbf{B}}\nc{\bbC}{\mathbf{C}}\nc{\bbD}{\mathbf{D}}
\nc{\bbE}{\mathbf{E}}\nc{\bbF}{\mathbf{F}}\nc{\bbG}{\mathbf{G}}\nc{\bbH}{\mathbf{H}}
\nc{\bbI}{\mathbf{I}}\nc{\bbJ}{\mathbf{J}}\nc{\bbK}{\mathbf{K}}\nc{\bbL}{\mathbf{L}}
\nc{\bbM}{\mathbf{M}}\nc{\bbN}{\mathbf{N}}\nc{\bbO}{\mathbf{O}}\nc{\bbP}{\mathbf{P}}
\nc{\bbQ}{\mathbf{Q}}\nc{\bbR}{\mathbf{R}}\nc{\bbS}{\mathbf{S}}\nc{\bbT}{\mathbf{T}}
\nc{\bbU}{\mathbf{U}}\nc{\bbV}{\mathbf{V}}\nc{\bbW}{\mathbf{W}}\nc{\bfX}{\mathbf{X}}
\nc{\bbY}{\mathbf{Y}}\nc{\bbZ}{\mathbf{Z}}

\nc{\sA}{\mathsf{A}}\nc{\sB}{\mathsf{B}}\nc{\sC}{\mathsf{C}}\nc{\sD}{\mathsf{D}}
\nc{\sE}{\mathsf{E}}\nc{\sF}{\mathsf{F}}\nc{\sG}{\mathsf{G}}\nc{\sH}{\mathsf{H}}
\nc{\sI}{\mathsf{I}}\nc{\sJ}{\mathsf{J}}\nc{\sK}{\mathsf{K}}\nc{\sL}{\mathsf{L}}
\nc{\sM}{\mathsf{M}}\nc{\sN}{\mathsf{N}}\nc{\sO}{\mathsf{O}}\nc{\sP}{\mathsf{P}}
\nc{\sQ}{\mathsf{Q}}\nc{\sR}{\mathsf{R}}\nc{\sS}{\mathsf{S}}\nc{\sT}{\mathsf{T}}
\nc{\sU}{\mathsf{U}}\nc{\sV}{\mathsf{V}}\nc{\sW}{\mathsf{W}}\nc{\sX}{\mathsf{X}}
\nc{\sY}{\mathsf{Y}}\nc{\sZ}{\mathsf{Z}}

\newcommand{\mathset}[1]{\left\{#1\right\}}
\newcommand{\multiset}[1]{\left\{\left\{#1\right\}\right\}}
\newcommand{\abs}[1]{\left|#1\right|}
\newcommand{\ceilenv}[1]{\left\lceil #1 \right\rceil}
\newcommand{\floorenv}[1]{\left\lfloor #1 \right\rfloor}
\newcommand{\parenv}[1]{\left( #1 \right)}
\newcommand{\sparenv}[1]{\left[ #1 \right]}
\newcommand{\bracenv}[1]{\left\{ #1 \right\}}
\nc{\set}[1]{\llbracket #1 \rrbracket}

\newcommand{\bal}[1]{\begin{align}\label{#1}}
\newcommand{\eal}{\end{align}}

\renewcommand{\le}{\leqslant}
\renewcommand{\leq}{\leqslant}
\renewcommand{\ge}{\geqslant}
\renewcommand{\geq}{\geqslant}

\renewcommand{\Bbb}{\mathbb}

\newcommand{\Tref}[1]{Theo\-rem\,\ref{#1}}

\newcommand{\Lref}[1]{Lem\-ma\,\ref{#1}}

\newcommand{\Examref}[1]{Example\,\ref{#1}}
\newcommand{\Obsref}[1]{Observation\,\ref{#1}}
\newcommand{\Tabref}[1]{Table\,\ref{#1}}
\newcommand{\Remarkref}[1]{Remark\,\ref{#1}}
\newcommand{\Dref}[1]{Definition\,\ref{#1}}

\renewcommand{\Bbb}{\mathbb}

\newcommand{\N}{{\Bbb N}}

\newcommand{\Z}{{\Bbb Z}}

\nc{\vt}{\vartheta}
\nc{\vtp}{\vartheta_{\bbP}}
\nc{\vtpk}{\vartheta_{\bbP,k}}

\outer\def\proclaim #1. #2\par{\medbreak
 \noindent{\bf#1.\enspace}{\sl#2\par}%
 \ifdim\lastskip<\medskipamount \removelastskip\penalty55\medskip\fi}

\begin{document}

\title{Codes Over Absorption Channels}

\author{\IEEEauthorblockN{Zuo~Ye and Ohad ~Elishco}
}

\maketitle

{\renewcommand{\thefootnote}{}\footnotetext{

\vspace{-.2in}
 
\noindent\rule{1.5in}{.4pt}

{ The authors are with the School of Electrical and Computer Engineering, Ben-Gurion University of the Negev, 
Beer Sheva, Israel. Email: \{zuoy,ohadeli\}@bgu.ac.il. 
}
}}



\begin{abstract} 
In this paper, we present a novel communication channel, called the absorption channel, inspired by information transmission in neurons. Our motivation comes from in-vivo nano-machines, emerging medical applications, and brain-machine interfaces that communicate over the nervous system. Another motivation comes from viewing our model as a specific deletion channel, which may provide a new perspective and ideas to study the general deletion channel. 

For any given finite alphabet, we give codes that can correct absorption errors. For the binary alphabet, the problem is relatively trivial and we can apply binary (multiple-) deletion correcting codes. For single-absorption error, we prove that the Varshamov-Tenengolts codes can provide a near-optimal code in our setting. When the alphabet size $q$ is at least $3$, we first construct a single-absorption correcting code whose redundancy is at most $3\log_q(n)+O(1)$. Then, based on this code and ideas introduced in \cite{Gabrys2022IT}, we give a second construction of single-absorption correcting codes with redundancy $\log_q(n)+12\log_q\log_q(n)+O(1)$, which is optimal up to an $O\left(\log_q\log_q(n)\right)$.

Finally, we apply the syndrome compression technique with pre-coding to obtain a subcode of the single-absorption correcting code. This subcode can combat multiple-absorption errors and has low redundancy. For each setup, efficient encoders and decoders are provided.
\end{abstract}

\section{Introduction}
The field of molecular or chemical communication, which involves the use of chemical signals for communication, has gained popularity in recent years due to advances in nano-technology and the development of nano-machines. These small devices can perform various tasks such as computing, storing data, transmitting information, and measuring physical quantities, and can be connected together to form a nano-network. Nano-networks are expected to have significant potential in future medical technologies, such as being used as an effective drug delivery system \cite{vashist2005smart,Davis1997} or for detecting infections through monitoring the values of different molecules \cite{dubach2007fluorescent,li2003cholesterol,tallury2010nanobioimaging,akan2016fundamentals}.

However, the small size of nano-machines presents challenges for traditional forms of communication \cite{yang2020comprehensive}, leading to the development of chemical communication as an alternative \cite{Akyildiz2008,farsad2016comprehensive}. This allows nano-machines to directly communicate with and across the human nervous system using chemical signals \cite{pan2022molecular,Chen2011,malak2012molecular}. There have been several communication models proposed and studied in this field \cite{gerstner2002spiking,malak2012molecular,malak2014communication,akan2016fundamentals}, and in one practical application, researchers transferred information through an in-vivo nervous system and observed the response of nerves to different voltages and frequencies \cite{abbasi2018controlled}.

In this paper, we propose a new type of transmission channel called \textbf{absorption channels}, which are inspired by neural and chemical communication systems. Our goal is to model a communication channel between nano-machines located within a living organism that utilize the organism's nervous system for communication and data collection. While chemical communication systems have been analyzed from an information-theoretic perspective, no coding-theoretic framework has been proposed. Therefore, the models we present in this paper are adapted to a coding-theoretic framework and are analyzed from a coding-theoretic perspective. 

An absorption error can be defined as follows: given a finite alphabet $\Sigma_q=\mathset{0,1,\dots, q-1}$ and an $n$-length sequence $\bfx=x_1 x_2 \dots x_n\in\Sigma_q^n$, the transmission of $\bfx$ through a single-absorption channel (which results in a single absorption error) produces an $(n-1)$-length sequence $x_1 \dots x_{i-1} (x_i\oplus x_{i+1}) x_{i+2} \dots x_n\in\Sigma_q^{n-1}$ for some $1\leq i\leq n-1$, where $a\oplus b\triangleq\min\mathset{a+b,q-1}$. 

To better demonstrate the connection between absorption channels and neural communication channels, we provide a brief explanation of neuron activity (for a more detailed explanation of neurons, see \cite[Ch. 8-11]{Silv}). Every cell, including nerve cells, consists of a fluid and particles encased in a membrane that allows certain materials and particles to pass through for communication with the surrounding environment. Neurons, or nerve cells, have several parts: dendrites, cell body, axon, and axon terminals (as shown in Figure 1). The dendrites are thin, branching extensions of the cell body that receive signals from other cells, the cell body contains the nucleus and other organelles, the axon is a long, thin projection that carries signals away from the cell body, and the axon terminals are the ending points of the axon that transmit signals to other cells. 

\begin{figure}[!htbp]
    \centering
    \includegraphics[width=0.6\columnwidth,height=0.23\linewidth]{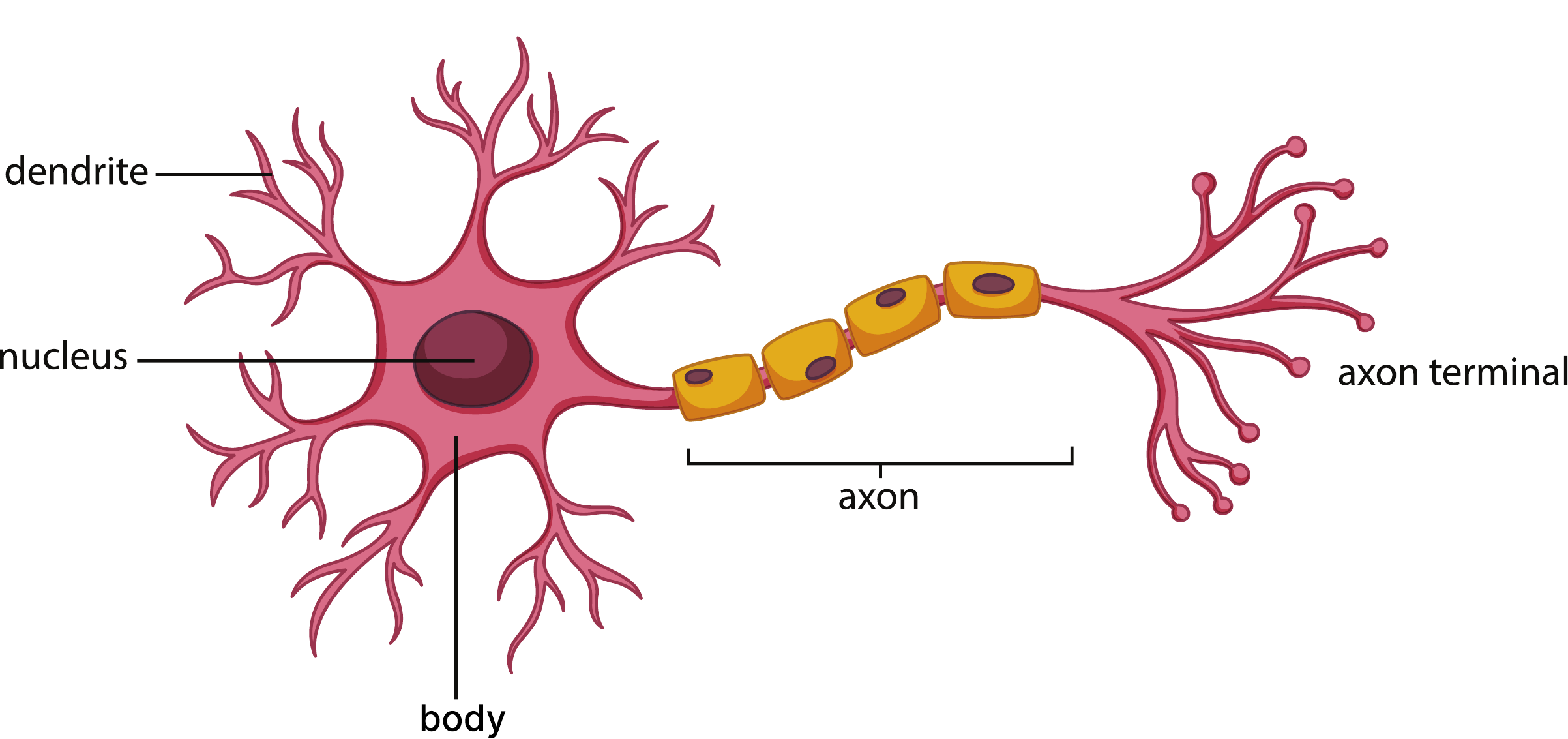}
    \caption{An illustration of a neuron (nerve cell) with its different parts: the dendrites, the cell body which contains the nucleus), the axon, and the axon terminals (downloaded from Vecteezy.com).}
    \label{fig_nerve}
\end{figure}

Neurons are specialized cells that transmit electrical and chemical signals within the nervous system (see \Cref{fig_nerve} for an illustration). 
To transmit a signal, a neuron generates an electrical charge, known as an action potential, which travels along the surface of the cell. Action potentials typically begin at the dendrites of a neuron. When a neuron receives input from another neuron, it may trigger an action potential, which is generated by the movement of ions across the cell membrane. Once triggered, the action potential travels down the length of the neuron, passing through the cell body and axon, to the axon terminal. In response, the axon terminal releases chemical signals, called neurotransmitters, which bind to receptors on the dendrites of neighboring neurons. This transmission of the action potential from one neuron to another allows for the communication of information within the nervous system. 

However, in some cases, an action potential may not be reached even if the neuron is depolarized by neurotransmitters. For example, if the neurotransmitters do not bind to enough receptors, the potential of the cell may increase, but not to the level required to trigger an action potential. This phenomenon is known as subthreshold stimulus\footnote{In most mammals, the resting potential of a neuron is -70mV. This refers to the electrical potential across the cell membrane of the neuron when it is not actively transmitting an action potential. In order to fire, or transmit an action potential, a neuron must reach a potential of -50 mV. If this threshold is reached, the neuron will undergo a series of changes in ion concentrations that result in the rapid depolarization of the cell membrane. During this process, the potential of the cell increases to +30 mV before returning to the resting potential of -70 mV. This rapid change in potential, known as the action potential (or firing), allows for the transmission of information within the nervous system. If a neuron does not reach a potential of -50 mV, it will not fire an action potential. In this case, the neuron may be more excitable for a period of time after the failed attempt. This phenomenon is known as post-inhibitory rebound (see \cite{kandel2000principles,Silv}).}.

In order for nano-machines to use nerve cells as communication channels, one machine should release neurotransmitters at the dendrites of a nerve cell, and another machine should detect the release of neurotransmitters from the nerve cell. The amount of neurotransmitters released can be used to represent symbols, such as small, medium, and large amounts representing 0, 1, and 2, respectively\footnote{Neurons are not found individually, but rather as a group or tissue. In order to utilize the communication capabilities of a neuron, it is necessary to isolate a single neuron from the tissue and use it as a standalone communication channel. If an entire (healthy) tissue is activated, it can result in unintended changes or effects on the body.}. 

Neurotransmitters are chemical messengers that are produced within the cell body and then transported (by motor proteins) to the axon terminals, where they are stored until they are released in response to an action potential. However, there may be a shortage of neurotransmitters at the axon terminals due to their transport from the cell body, which can lead to a deficiency in the amount of neurotransmitters released when a neuron repeatedly fires \cite{Liu2014}. This deficiency can result in the transmission of a different ("lower valued") symbol. Additionally, the production rate and quantity of neurotransmitters is influenced, among other things, by the depolarization of the cell, and an excess of neurotransmitters at the axon terminals may lead to the release of an excess amount and the transmission of a different ("higher") symbol. 

The errors discussed above can occur in the context of communication between nano-machines using the nervous system as a transmission channel. If a symbol is to be transmitted while there is a deficiency of neurotransmitters, less neurotransmitters will be emitted and a "lower valued" symbol will be read. As a response to the deficiency, the neuron manufactures additional neurotransmitters. Thus, for the next transmission, an excess amount of neurotransmitters will be emitted and a "higher valued" symbol will be read. Similarly, if a transmission attempt depolarized the cell but not enough to reach an action potential, no neurotransmitters will be emitted (this corresponds to a deletion of the transmitted symbol). As a response to the depolarization, an additional amount of neurotransmitters is manufactured. Thus, in the next transmission attempt, an excess amount of neurotransmitters will be emitted and a "higher valued" symbol will be read. In this paper, we chose to focus only on the second error, in which a symbol is deleted and its value is added to the next transmission. 

Mathematically, these observations give rise to a family of communication channels. Let us consider the transmission of a string $\bfx \in \Sigma_q^n$ and the received string $\bby$. An error in the $i$th position can be described as follows: if the value of $y_i$ is smaller ($y_i < x_i$), then the missing value is added to the next symbol, meaning $y_{i+1} = \min\left(q-1, x_{i+1} + (y_i - x_i)\right)$. Alternatively, the $i$th symbol may be deleted completely ($\bby \in \Sigma_q^{n-1}$) and its value added to the next symbol, so $y_i = \min\left(q-1, x_i + x_{i+1}\right)$. In this work, we focus on the simplified case in which only the second error may occur, namely, the symbol is deleted and its entire value is added to the next symbol.

In addition to being motivated by neural communication systems, a single-absorption error can also be viewed as a deletion error followed by at most one substitution error. The study of codes that correct single-deletion and single-substitution errors was first introduced in the context of DNA-based data storage in \cite{hec2019} and further developed in \cite{Yaakobi2020isit}. More recently, codes that correct multiple-deletion and multiple-substitution errors were proposed in \cite{Song2022IT}. These results apply to our error model as well, but in this paper we demonstrate that it is possible to use specific absorption properties to achieve higher rates in our codes.

We also consider a variant of absorption errors called contraction errors, which we show are equivalent to deletion errors. The problem of constructing deletion-correcting codes dates back at least to the 1960s \cite{VL1966}. Recently, there has been renewed interest in this problem due to its potential applications in DNA-based data storage \cite{Sima2021it,Guruswami2021it} and document exchange \cite{Cheng2018FOCS,Haeupler2019FOCS}. Despite significant progress, constructing deletion-correcting codes remains a challenging problem with no complete solution. Our new findings may provide new insights into this problem.

The paper is organized as follows: in Section \ref{sec:pre}, we introduce the notation and definitions that will be used throughout the paper. In Section \ref{sec_binary}, we present codes over the binary alphabet. Section \ref{sec_nonbinary} contains the main results of this paper, which is the construction of absorption error-correcting codes for general alphabets. In Section \ref{sec_optimality}, we show that our single-absorption codes are asymptotically optimal in terms of redundancy. In Section \ref{sec_variant}, we study contraction errors as a variant of absorption errors and show that they are equivalent to deletion errors. Finally, in Section \ref{sec_conc}, we conclude the paper.

\section{Preliminary}\label{sec:pre}
For positive integers $m\le n$, let $[m,n]$ denote the set $\{m,m+1,\ldots,n\}$ and $[n]=\{1,\ldots,n\}$. 
For an integer $q\ge 2$, let $\Sigma_q$ denote the $q$-ary alphabet $\{0,1,\ldots,q-1\}$ and $\Sigma_q^n$ denote the set consisting of all length-$n$ sequences over $\Sigma_q$.
For any sequence $\bfx\in\Sigma_q^n$, unless otherwise stated, we let $x_i$ be the $i$th component of $\bfx$. In other words, $\bfx=x_1\cdots x_n$.
Suppose that two positive integers $n$ and $n^{\prime}$ satisfy $n\ge n^{\prime}$. Let $\bfx\in\Sigma_q^n$ and $\bby\in\Sigma_q^{n^{\prime}}$. If there are integers $1\le i_1<i_2<\cdots <i_{n^{\prime}}\le n$ such that $y_j=x_{i_j}$ for each $1\le j\le n^{\prime}$, we say that $\bby$ is a \emph{subsequence} of $\bfx$. If $I=\mathset{i_1,i_2,\ldots,i_{n^{\prime}}}$ (keep the order of $i_1,i_2,\ldots,i_{n^{\prime}}$), we also denote this subsequence by $\bfx_{I}$. Furthermore, if $i_{j+1}=i_j+1$ for all $1\le j<n^{\prime}$, we call $\bby$ a \emph{substring} of $\bfx$. A \emph{run} of $\bfx$ is a maximal substring consisting of identical symbols from $\Sigma_q$. If a run consists of symbol $a$, we say it is an \emph{$a$-run}. In this paper, the length of a sequence $\bfx$ is denoted by $\abs{\bfx}$.
\begin{example}
    Let $\bfx=001112\in\Sigma_3^6$, $\bby=012$ and $\bbz=0111$. Then $\bby$ is a subsequence of $\bfx$ and $\bbz$ is a substring of $\bfx$. Specifically, we have $\bby=\bfx_I$ and $\bbz=\bfx_J$, where $I=\mathset{1,3,6}$ and $J=\mathset{1,3,4,5}$. There are exactly three runs in $\bfx$: $00$, $111$ and $2$. They are $0$-run, $1$-run and $2$-run, respectively.
\end{example}

For $a,b\in\Sigma_q$, we define $a\oplus b=\min\{a+b,q-1\}$. Notice that $\oplus$ is an associative operation thus the order in which it is performed does not affect the result. 
Suppose $\bfx\in\Sigma_q^n$. We say that the sequence $\bby\in\Sigma_q^{n-1}$ is obtained from $\bfx$ by an \emph{absorption} if $\bby$ is either one of the following two cases:
\begin{enumerate}[$(1)$]
  \item $\bby=x_1\cdots x_{i-1}(x_i\oplus x_{i+1})x_{i+2}\cdots x_n$ for some $1\le i\le n-1$;
  \item $\bby=x_1\cdots x_{n-1}$.
\end{enumerate}
When the second case happens, we say that $x_n$ is \textit{missing}. Otherwise, we say that $x_n$ is \textit{not missing}.

For multiple absorptions, the situation becomes a little more complicated. For example, let $\bfx\in\Sigma_q^n$, then $\bby_1=x_1\cdots x_{i-1}(x_i\oplus x_{i+1})x_{i+2}\cdots x_{j-1}(x_j\oplus x_{j+1})x_{j+2}\cdots x_n$ where $i+2\le j<n$, and $\bby_2=x_1\cdots x_{i-1}(x_i\oplus x_{i+1})x_{i+2}\cdots x_{n-1}$ where $i<n-1$ are both obtained from $\bfx$ by two absorptions. 
Now let $\bby_3=x_1\cdots x_{i-1}(x_i\oplus x_{i+1}\oplus x_{i+2})x_{i+3}\cdots x_n$ where $i<n-1$. It is clear that $\bby_3$ can be obtained from $\bfx$ by first absorbing $x_i$ and $x_{i+1}$, and then absorbing $(x_{i}\oplus x_{i+1})$ and $x_{i+2}$.\footnote{or by first absorbing $x_{i+1}$ and $x_{i+2}$, and then absorbing $x_{i}$ and $(x_{i+1}\oplus x_{i+2})$.} Therefore, the sequence $\bby_3$ is also obtained from $\bfx$ by two absorptions. In general, we have the following definition.
\begin{definition}\label{dfn_absorption}
  Let $\bfx\in\Sigma_q^n$ and $\bby\in\Sigma_q^{n-t}$ where $n>t\ge1$. We say that $\bby$ is obtained from $\bfx$ by $t$ absorptions, if there is an integer $t^{\prime}\in\sparenv{0,t}$ and positive integers $k$, $s_l$ ($1\le l\le k$) and $i_l$ ($1\le l\le k$) satisfying $i_{l+1}-i_l>s_l$ for all $1\le l<k$ such that $\sum_{l=1}^{k}s_l=t-t^{\prime}$, $i_k+s_k\le n-t^{\prime}$ and
  \small{
  \begin{equation*}
y_i=
  \begin{cases}
    x_i, & \mbox{if }  i<i_{1},\\
    x_{i+\sum_{j=1}^{l}s_j}, & \begin{array}{l}\mbox{if } i_l-\sum_{j=1}^{l-1}s_j<i<i_{l+1}-\sum_{j=1}^{l}s_j\\
    \text{ for }1\le l<k,\end{array}\\
    x_{i+\sum_{j=1}^{k}s_j}, & \mbox{if } i>i_k-\sum_{j=1}^{k-1}s_j,\\
    \mathop{\bigoplus}\limits_{j=i_l}^{i_l+s_l}x_j, & \mbox{if } i=i_l-\sum_{j=1}^{l-1}s_j\text{ for }1\le l\le k.
  \end{cases}
\end{equation*}
}
Here, the substring $\bfx_{\sparenv{n-t^{\prime}+1,n}}$ is deleted.
\end{definition} 

In Definition \ref{dfn_absorption}, the starting positions of the absorptions are denoted by the $i_l$s, while the number of symbols absorbed with $x_{i_l}$ is denoted by $s_l$.

\begin{example}
  Let $\Sigma_3$ be the ternary alphabet and let $\bfx=011011111$. Assume there are $t=3$ absorptions, with $t^{\prime}=0$, $k=2$, $s_1=2,\; s_2=1$ and $i_1=2,\; i_2=6$. 
  The resulting sequence is $\bby_1=021211$. 
  
  Now assume $t^{\prime}=1$, $k=1$, $s_1=2$, and $i_1=2$, then $\bby_2=021111$ with $x_n$ missing.
\end{example}

For a sequence $\bfx\in\Sigma_q^n$ and a positive integer $t$ satisfying $t<n$, we define the set
\begin{equation}\label{eq_absorptionball}
  \cB_t^{ab}\parenv{\bfx}\triangleq\mathset{\bby\in\Sigma_q^{n-t}~:~\bby\text{ is obtained from }\bfx\text{ by }t\text{ absorptions}}
\end{equation}
and call it the \emph{$t$-absorption ball} centered at $\bfx$. Note that $\cB_t^{ab}\parenv{\bfx}$ depends on the alphabet $\Sigma_q$. We omit $q$ in this notation since the alphabet will be clear from the context.
\begin{definition}
    Let $t$ be a positive integer. Let $\cC$ be a nonempty subset of $\Sigma_q^n$. If $\cB_t^{ab}\parenv{\bfx}\cap\cB_t^{ab}\parenv{\bby}=\emptyset$ for any distinct $\bfx,\bby\in\cC$, we call it a \emph{$t$-absorption correcting code}. The \emph{redundancy} of $\cC$ is defined to be $n-\log_q\parenv{\abs{\cC}}$. In other words, the redundancy is measured in $q$-ary symbols.
\end{definition}

In this paper, we aim to construct $t$-absorption correcting codes with low redundancy, for any $t$. Throughout this paper, the number of errors $t$ and the alphabet size $q$ are assumed to be fixed constants.

\section{Codes over Binary alphabet}\label{sec_binary}
In this section we present a construction of a binary code that can repair multiple absorptions. The construction relies on the following simple observation. 

\begin{observation}\label{obs_binary}
Suppose that $\bby$ is obtained from $\bfx\in\Sigma_2^n$ by absorbing $x_i$ and $x_{i+1}$. If $x_ix_{i+1}\in\{00,01,10\}$, then $\bby$ is obtained from $\bfx$ by deleting one $0$. If $x_ix_{i+1}=11$, then $\bby$ is obtained from $\bfx$ by deleting one $1$. Therefore, no matter whether the last symbol $x_n$ is lost or not, $\bby$ is obtained from $\bfx$ by deleting one symbol.
\end{observation}
Notice that when at most one absorption occurs, an isolated $1$ cannot be deleted, i.e., any $1$ that both of its neighbors are $0$, will not be deleted. 

From \Obsref{obs_binary}, we have that every binary single-deletion correcting code is a binary single-absorption correcting code. The opposite, however, is not necessarily true, as shown in the next example. 

\begin{example}\label{examp_notequiv}
We give an example to show that a single-absorption correcting code is not necessarily a single-deletion correcting code.
Consider the code $\mathset{011000,011010}$. A single absorption on $011000$ yields 3 possible outputs as before: $11000,01000,01100$. A single absorption on $011010$ yields also 3 possible outputs: $11010,01010,01110$. However, the sequence $01100$ can be obtained from both codewords by deleting the one-before-last symbol. Thus, the code cannot correct a single deletion.
\end{example}

To construct a single-absorption correcting code, 
we can use \Obsref{obs_binary} and apply single-deletion correcting codes. The best-known class of binary single-deletion correcting codes are the famous Varshamov-Tenengolts (VT) codes \cite{VT1965}, which are defined as
\begin{align}\label{eq_vtcode}
  \mathrm{VT}_a(n)=\left\{\bbc\in\Sigma_2^n~:~\Syn{\bbc}\equiv a\pmod{n+1}\right\},
\end{align}
where $a$ is an integer between $0$ and $n$, and $\Syn{\bbc}\triangleq\sum_{i=1}^{n}ic_i$ is the VT \emph{syndrome} of $\bm{c}$. The smallest redundancy of $\log_2(n+1)$ is attained when $a=0$ \cite[Corollary 2.3]{Sloane2002}. A linear-time decoding algorithm of the VT codes to correct a single deletion was provided in \cite{VL1966}. In \cite{GhaFer1998} the authors gave a linear-time systematic encoder with redundancy $\ceilenv{ \log_2(n+1)}$. 


\begin{remark}
  By \Examref{examp_notequiv}, one can deem that there might be a single-absorption correcting code of length $n$ with a larger size than that of the VT code $\mathrm{VT}_0(n)$. In \Cref{sec_optimality}, we will show that for single-absorption, the redundancy of the code $\mathrm{VT}_0(n)$ is optimal up to a constant. 
\end{remark}

By \Dref{dfn_absorption}, it is not difficult to see that \Obsref{obs_binary} can be generalized to the case when multiple absorptions happen. 
To be specific, if $\bby$ is obtained from $\bfx$ by $t$ absorptions, then it is obtained from $\bfx$ by $t$ deletions. So we can apply multiple-deletion correcting codes for our setting. There are already a myriad of works on binary multiple-deletion correcting codes (see, for example, \cite{Joshua2018it,Gabrys2019it,Sima2021it,Sima2020it,Guruswami2021it,Song2022IT}). For $t=2$, the best known result was given in \cite{Guruswami2021it}, where an explicit binary $2$-deletion correcting code of length $n$ with redundancy at most $4\log_2(n)+O(\log_2\log_2(n))$ was constructed. This code is polynomial-time encodable and decodable. For general $t\ge3$, the best known result was contributed in \cite{Song2022IT}, where the authors proved that there is a binary systematic $t$-deletion correcting code of length $n$ with redundancy at most $(4t-1)\log_2(n)+o(\log_2(n))$. The encoding and decoding complexities are $O\parenv{n^{2t+1}}$ and $O\parenv{n^{t+1}}$ respectively.

\section{Codes over non-binary alphabets} \label{sec_nonbinary}
In \Cref{sec_binary}, we showed that a single-absorption error is a special case of single-deletion error. The situation is different when the alphabet size is at least $3$. Throughout this section, we always assume that the alphabet is $\Sigma_q$, where $q\ge 3$. 
This section contains three parts. At first, we present a basic code construction that can correct a single absorption. In the second part, we improve upon the basic construction and present a construction with smaller redundancy that can correct a single absorption error. 
In the last part, we study codes that can correct multiple absorptions. 

\subsection{A basic construction}
We begin with a construction of a single absorption correcting code.

For a sequence $\bfx\in\Sigma_q^n$ and a symbol $a\in\Sigma_q$, we let $N_a(\bfx)$ to be the number of $a$ appearing in $\bfx$, that is,
\begin{equation*}
  N_a(\bfx)\triangleq\left|\left\{i:\;x_i=a\right\}\right|.
\end{equation*}
Let $\bby=x_1\cdots x_{i-1}(x_i\oplus x_{i+1})x_{i+2}\cdots x_n$ be the received sequence, where $1\le i\le n-1$\footnote{As will be clear later, our code can correct a single deletion. So we do not need to discuss the case $\bby=\bfx_{[1,n-1]}$}.

\begin{observation}\label{obs_general}
Let $a,b\in\Sigma_q$ and $0<a,b< q-1$.
\begin{itemize}
  \item If $x_ix_{i+1}\in\{0a,a0,00\}$, then $N_0(\bfx)=N_0(\bby)+1$ and $N_d(\bfx)=N_d(\bby)$ for all $d\ne 0$. In other words, $\bby$ is obtained from $\bfx$ by deleting one $0$.
  \item If $x_i=x_{i+1}=q-1$, then $N_{q-1}(\bfx)=N_{q-1}(\bby)+1$ and $N_d(\bfx)=N_d(\bby)$ for all $d\ne q-1$. In other words, $\bby$ is obtained from $\bfx$ by deleting one $q-1$.
  \item If $x_ix_{i+1}\in\{(q-1)a,a(q-1)\}$, then $N_a(\bfx)=N_a(\bby)+1$ and $N_d(\bfx)=N_d(\bby)$ for all $d\ne a$. In other words, $\bby$ is obtained from $\bfx$ by deleting one $a$.
  \item If $x_ix_{i+1}=ab$ and $c=a\oplus b$, then $N_a(\bfx)=N_a(\bby)+1$, $N_b(\bfx)=N_b(\bby)+1$, $N_c(\bfx)=N_c(\bby)-1$ and $N_d(\bfx)=N_d(\bby)$ for all $d\ne a,b,c$ (if $a=b$ then $N_a(\bfx)=N_a(\bby)+2$).
\end{itemize}
\end{observation}

From \Obsref{obs_general}, we can see that if $0<x_i,x_{i+1}<q-1$, a single-absorption error is a single-deletion together with a single-substitution, which is different from the binary case. 
In general, a single-absorption error is a single-deletion together with at most a single-substitution (no matter whether the last symbol is missing or not). 
Therefore, if a code can combat a single-deletion together with at most a single-substitution, it can also correct a single-absorption error. 
In \cite{Yaakobi2020isit}, the study of single-deletion single-substitution codes was initiated, and the authors gave a $q$-ary single-deletion single-substitution correcting code of redundancy at most $10\log_2(n)+O(1)$ (measured in bits) \cite[Corollary 17]{Yaakobi2020isit}. For more details about this kind of codes, we refer the interested readers to \cite{Yaakobi2020isit} and \cite{Song2022IT}. 

At this point, one may wonder if a single-absorption error correcting code is also a single-deletion single-substitution error correcting code. To answer this, we first notice that the substitution caused by an absorption is specific and depends on the absorbed symbol. Thus, it is reasonable to assume that a single-absorption error is a specific case of a single-deletion single-substitution error. Indeed, as shown in the next example, a single-absorption correcting code is not necessarily a single-deletion single-substitution correcting code.
\begin{example}
  Let $q=3$ and consider the code $\mathset{110110,011010}$. A single absorption error on $110110$ yields one of the following words: $20110, 11110, 11020, 11011$; a single absorption error on $011010$ yields one of the following words: $11010, 02010, 01110, 01101$. Therefore, this code can correct a single-absorption error. On the other hand, the sequence $11010$ can be obtained from $110110$ and from $011010$ by a single deletion. So this code cannot correct a single-deletion and at most a single-substitution error.
\end{example}

Thus, one may infer that there might be codes with lower redundancy for absorption channels. In this section, we show that indeed it is possible to obtain codes with less redundancy.

We begin with constructing a set of $n$-length words over $\Sigma_q$, which is defined by a vector $\bm{s}$ of length $(q-1)$ over $\Z_4$. 
Given $\bm{s}=(s_a)_{a\in[0,q-2]}\in\mathbb{Z}_4^{q-1}$, we define
\begin{equation*}
  \mathcal{C}_1(n;\bm{s})\triangleq\left\{\bfx\in\Sigma_q^n:\;N_{a}(\bfx)\equiv s_a\pmod{4}\text{ for each }a\in[0,q-2]\right\}.
\end{equation*}
Since $N_{q-1}(\bfx)=n-\sum_{a=0}^{q-2}N_a(\bfx)$, we can obtain $N_{q-1}(\bfx)\pmod{4}$ when given all $N_{a}(\bfx)\pmod{4}$ where $a\in[0,q-2]$. 
Assume a single absorption channel, and suppose that a transmitted sequence $\bfx$ is in $\mathcal{C}_1(n;\bm{s})$. Denote the obtained sequence (the channel output) by 
$\bby$. 
Since $\bby$ is obtained from $\bfx$ by a single absorption, this absorption must be one of the four cases described in \Obsref{obs_general}.
By calculating $N_a(\bby)-s_a\pmod{4}$ for all $a\in\Sigma_q$,\footnote{where $s_{q-1}\in\mathset{0,1,2,3}$ and satisfies $s_{q-1}\equiv n-\sum_{a=0}^{q-2}s_a\pmod{4}$} it is possible to know which one of the cases happened (without knowing the position in which the absorption happened). The details are shown in \Tabref{tab_cases}. Hereafter, let $\multiset{\cdot}$ denote a multiset.
\begin{table}[!htbp]
	\centering
	\caption{The relation between $N_a(\bby)-s_a$ ($a\in\Sigma_q$) and the values of $x_i$ and $x_{i+1}$}
	\label{tab_cases}
	\begin{tabular}{c|c}
		\hline
		\hline
		Cases& The values of $x_i$ and $x_{i+1}$\\
		\hline
		\tabincell{l}{$N_0(\bby)-s_0\equiv 3\pmod{4}$ and\\ $N_a(\bby)-s_a\equiv 0\pmod{4}$ for all $a\ne0$}&$0\in\multiset{x_i,x_{i+1}}$\\
		\hline
	\tabincell{l}{$N_a(\bby)-s_a\equiv 3\pmod{4}$ for some $a\ne0$ and\\ $N_b(\bby)-s_b\equiv 0\pmod{4}$ for all $b\ne a$}&$\multiset{x_i,x_{i+1}}=\multiset{a,q-1}$\\
	\hline
	\tabincell{l}{$N_a(\bby)-s_a\equiv 2\pmod{4}$,\\ $N_c(\bby)-s_c\equiv 1\pmod{4}$ for some $a,c$, and\\
	$N_b(\bby)-s_b\equiv 0\pmod{4}$ for all $b\ne a,c$}&\tabincell{c}{$x_i=x_{i+1}=a$\\$0<a<q-1$}\\
\hline
\tabincell{l}{$N_a(\bby)-s_a\equiv 3\pmod{4}$,\\
	$N_b(\bby)-s_b\equiv 3\pmod{4}$,\\
	$N_c(\bby)-s_c\equiv 1\pmod{4}$ for some $a,b,c$, and\\
	$N_d(\bby)-s_d\equiv 0\pmod{4}$ for all $d\ne a,b,c$}&\tabincell{c}{$\mathset{x_i,x_{i+1}}=\mathset{a,b}$\\$0<a,b<q-1$\\$a\ne b$}\\
		\hline
	\end{tabular}
\end{table}

Thus, if a sequence $\bfx\in\cC_1(n;\bm{s})$ is transmitted through a single absorption channel and $\bby$ is the output of the channel, it is possible to distinguish which one of the four absorption cases described in \Obsref{obs_general} has occurred. 
However, more information is needed in order to recover $\bfx$ from $\bby$. For example, the order of the absorbed symbols (if $a\oplus b=c$ then also $b\oplus a=c$), or the exact position of the absorption.  
Therefore, we need to add additional redundancy layers to $\cC_1(n;\bm{s})$ as explained next. 

To account for the order of the absorbed symbols, let us first consider the case $x_ix_{i+1}\in\multiset{ab,ba}$ where $0<a,b<q-1$ and $a,b$ are not necessarily distinct. As mentioned above, by calculating $N_d(\bby)-s_d\pmod{4}$ for all $d\in\Sigma_q$, one can deduce the values of $a,b$ and $c=a\oplus b$, but cannot necessarily deduce their order ($ab$ or $ba$). 
In order to distinguish between the two cases $x_ix_{i+1}=ab$ or $x_ix_{i+1}=ba$, we need the following notation: for any $\bbz\in\Sigma_q^n$, let
\begin{equation*}
  \Inv{\bbz}\triangleq\left|\left\{(i,j):\;1\le i<j\le n,z_i>z_j\right\}\right|.
\end{equation*}
Let $\bfx^{\prime}$ and $\bfx^{\prime\prime}$ be the sequences obtained from $\bby$ by replacing a specific $c$ with $ab$ and $ba$, respectively. 
Then $\Inv{\bfx^{\prime}}-\Inv{\bfx^{\prime\prime}}=\pm 1$. Therefore, if we fix $\Inv{\bfx}\pmod{2}$ and this value is known, we obtain that at most one of $\bfx^{\prime}$ and $\bfx^{\prime\prime}$ equals $\bfx$. 

Now, consider the case when $0\in\multiset{x_i,x_{i+1}}$ or $q-1\in\multiset{x_i,x_{i+1}}$. In this case, $\bby$ is obtained from $\bfx$ by a single deletion. In order to correct such an error, we need a $q$-ary code that can correct a single deletion. 

For each $\bbz\in\Sigma_q^n$, let $\alpha(\bbz)\in\Sigma_2^{n-1}$, where $\alpha(\bbz)_i=1$ if $z_{i+1}\ge z_i$, and $0$ otherwise for each $i\in[n-1]$. For given $t_1\in\mathbb{Z}_{n}$ and $t_1^{\prime}\in\mathbb{Z}_q$, it was shown in \cite{Tenengolts1984} that the following $q$-ary code can correct a single deletion:
\begin{equation*}
  T_{t_1,t_1^{\prime}}(n;q)\triangleq\left\{\bbz\in\Sigma_q^n:\;\Syn{\alpha(\bbz)}\equiv t_1\pmod{n},\mathop{\sum}\limits_{i=1}^{n}z_i\equiv t_1^{\prime}\pmod{q}\right\}.
\end{equation*}
However, the only role of the constraint $\mathop{\sum}\limits_{i=1}^{n}z_i\equiv t_1^{\prime}\pmod{q}$ is to determine the deleted symbol. 
In our setting, the deleted symbol is known by calculating $N_a(\bby)-s_a\pmod{4}$, so we do not need this constraint (in fact, in our case this constraint is replaced with the constraint $N_a(\bfx)\equiv s_a\pmod{4}$).

Putting what we have so far together, we construct the following code. For a given 
$\bm{s}=(s_a)_{a\in[0,q-2]}\in\mathbb{Z}_4^{q-1}$ and $\bm{t}=(t_1,t_2)\in\mathbb{Z}_n\times\mathbb{Z}_2$, let
\begin{align*}
  \mathcal{C}_2(n;\bm{s},\bm{t})&\triangleq\left\{\bfx\in\mathcal{C}_1(n;\bm{s}):\;\Syn{\alpha(\bbz)}\equiv t_1\pmod{n},\Inv{\bfx}\equiv t_2\pmod{2}\right\}.
\end{align*}
Let $\bfx\in\mathcal{C}_2(n;\bm{s},\bm{t})$ and let $\bby$ be the sequence received after transmitting $\bfx$ through a single-absorption channel. 
By the discussions above, if $0\in\multiset{x_i,x_{i+1}}$ or $q-1\in\multiset{x_i,x_{i+1}}$, we can recover $\bfx$ from $\bby$ by the decoder of $T_{t_1,t_1^{\prime}}(n;q)$. 
If $x_ix_{i+1}\in\multiset{ab,ba}$ where $0<a,b<q-1$, we can find the values of $a$ and $b$ using $\bm{s}$ and \Tabref{tab_cases}, and for the specific $c=a\oplus b$ in $\bby$ that was obtained by the absorption, we can determine whether $x_ix_{i+1}=ab$ or $x_ix_{i+1}=ba$ using $\Inv{\bfx}$. What we are still missing in order to be able to recover $\bfx$ is the exact absorption position, i.e., the position of that $c=a\oplus b$. 

Our next aim is to add another layer of redundancy that determines the position of absorption in the case that  $x_ix_{i+1}\in\multiset{ab,ba}$ with $0<a,b<q-1$. 
We will divide our discuss into two cases. Different methods will be applied to locate the error position.

\noindent (1) \textbf{The Case $a+b\le q-1$}

Let $\bfx^{\prime}$ be the sequence obtained from $\bby$ by replacing the $c$ located at position $i$ with one of $ab$ and $ba$, $\bfx^{\prime\prime}$ be the sequence obtained from $\bby$ by replacing the $c$ located at position $j$ with one of $ab$ and $ba$, where $1\le i<j\le n-1$. Recall that $\Syn{\bbz}=\sum_{i=1}^{\abs{\bbz}}iz_i$ for any sequence $\bbz$.
\begin{lemma}\label{lem_smallvalue}
    $\Syn{\bfx^{\prime}}\not\equiv\Syn{\bfx^{\prime\prime}}\pmod{qn}$.
\end{lemma}
\begin{IEEEproof}
   Since $a+b\le q-1$, we have $a\oplus b=a+b$. Then it is easy to see that 
    \begin{align*}
    \Syn{\bfx^{\prime}}-\Syn{\bby}=\alpha+\sum_{k=i+1}^{n-1}y_k,\\
    \Syn{\bfx^{\prime\prime}}-\Syn{\bby}=\beta+\sum_{k=j+1}^{n-1}y_k,
    \end{align*}
    where $\alpha,\beta\in\multiset{a,b}$. These two equations imply that
    $$
    \Syn{\bfx^{\prime}}-\Syn{\bfx^{\prime\prime}}=\alpha-\beta+\sum_{k=i+1}^{j}y_k=
    \begin{cases}
        y_j+\sum_{k=i+1}^{j-1}y_k,&\mbox{ if }\alpha=\beta,\\
        2\alpha+\sum_{k=i+1}^{j-1}y_k,&\mbox{ if }\alpha\ne\beta.
    \end{cases}
    $$
    Noticing that $y_j=a+b$ and $a,b>0$, we have
    $$
   0<\Syn{\bfx^{\prime}}-\Syn{\bfx^{\prime\prime}}<qn.
    $$
    Now the proof is completed.
\end{IEEEproof}

By \Lref{lem_smallvalue}, if we fix  $\Syn{\bfx}\pmod{qn}$, where $\bfx\in\cC_2(n;\bm{s},\bm{t})$, then we can find a unique $c$ in $\bby$ such that $\bfx$ is obtained from $\bby$ by replacing this $c$ with $ab$ or $ba$. Details will be shown in the proof of \Cref{thm_primarycode} below.

\noindent (2) \textbf{The Case $a+b\ge q$}

In this case, we have $a\oplus b=q-1$. We want to locate in $\bby$ the position of the symbol $q-1$ which is obtained by a single absorption. To this end, we define the location sequence of a sequence $\bfx\in\Sigma_q^n$ to be $P\parenv{\bfx}\in\Sigma_2^n$, where
$$
P\parenv{\bfx}_i=
\begin{cases}
    0,&\mbox{ if }x_i\ne q-1,\\
    1,&\mbox{ if }x_i=q-1.
\end{cases}
$$
Suppose $\bby$ is obtained from $\bfx$ by absorbing $x_i$ and $x_{i+1}$, where $0<x_i,x_{i+1}<q-1$ and $x_i+x_{i+1}\ge q$. It is easy to see that $P\parenv{\bby}$ is obtained from $P\parenv{\bfx}$ by replacing two adjacent $0$s with a single $1$. We call this error type $00\rightarrow 1$. Now locating the error position in $\bfx$ is reduced to locating the error position in $P\parenv{\bfx}$. For this, we have the following code.

For any $n\ge 3$ and any $d\in\mathbb{Z}_{2n-3}$, define
\begin{equation*}
\cC_3\parenv{n;d}\triangleq\mathset{\bbz\in\Sigma_2^n~:~\Syn{\bbz}\equiv d\pmod{2n-3}}.
\end{equation*}
\begin{lemma}\label{lem_largevalue}
    The binary code $\cC_3\parenv{n;d}$ can correct the error type $00\rightarrow 1$ and locate the error position.
\end{lemma}
\begin{IEEEproof}
    Suppose that $\bbz^{\prime}$ is obtained from a codeword $\bbz\in\mathrm{VT}_d(2n-3)$ by the error $00\rightarrow1$. Let the two sequences $\bbu$ and $\bbv$ be obtained from $\bbz^{\prime}$ by replacing $z_i^{\prime}=1$ and $z_j^{\prime}=1$ with $00$, respectively, where $1\le i<j\le n-1$. Then $\Syn{\bbu}-\Syn{\bbv}=j-i+\sum_{k=i+1}^jz_k^{\prime}$. So we have
    \begin{equation}\label{eq_difference}
    0<j-i\le\Syn{\bbu}-\Syn{\bbv}\le 2(j-i)\le 2n-4<2n-3.
   \end{equation}
   Now we can recover $\bbz$ from $\bbz^{\prime}$ by the following procedure. Scan the symbols from the beginning of $\bbz^{\prime}$  to its end. If the symbol $1$ is encountered, conduct the following steps.
   \begin{enumerate}[\textbf{Step} 1]
   \item Replace this $1$ with $00$ and denote the resulting sequence by $\bbu$. If $\Syn{\bbu}\equiv d\pmod{2n-3}$, let $\bbz$ and output $\bbz$. Otherwise, go to Step 2.
   \item Move to the next $1$ and go to Step 1.
   \end{enumerate}
   Since $\bbz^{\prime}$ is obtained from $\bbz$ by the error type $00\rightarrow 1$, this $\bbu$ does exist. On the other hand, \Cref{eq_difference} ensures that such $\bbu$ is unique and the error position can be uniquely determined.
\end{IEEEproof}

Now we are ready to give a code that can correct a single-absorption error. Given $n\ge 3$, $\bm{s}=(s_a)_{a\in[0,q-2]}\in\mathbb{Z}_4^{q-1}$, $\bm{t}=(t_1,t_2)\in\mathbb{Z}_n\times\mathbb{Z}_{2}$ and $\bm{d}=\parenv{d_1,d_2}\in\mathbb{Z}_{qn}\times\mathbb{Z}_{2n-3}$, let
\begin{equation*}
  \mathcal{C}(n;\bm{s},\bm{t},\bm{d})
  \triangleq\mathset{\bfx\in\mathcal{C}_2(n;\bm{s},\bm{t})~:~\Syn{\bfx}\equiv d_1\pmod{qn},P\parenv{\bfx}\in\cC_3\parenv{n;d_2}}.
\end{equation*}

\begin{theorem}\label{thm_primarycode}
  The code $\cC\parenv{n;\bm{s},\bm{t},\bm{d}}$ can correct a single-absorption error.
\end{theorem}
\begin{IEEEproof}
  Let $\bfx\in\cC\parenv{n;\bm{s},\bm{t},\bm{d}}$ be the transmitted sequence and $\bby\in\Sigma_q^{n-1}$ be the received sequence. Suppose that $\bby$ is obtained from $\bfx$ by replacing $x_ix_{i+1}$ with $c=x_i\oplus x_{i+1}$. Since $\bfx\in\mathcal{C}_1(n;\bm{s})$, we know if $0\in\multiset{x_i,x_{i+1}}$ or $q-1\in\multiset{x_i,x_{i+1}}$ or neither of the cases. If $0\in\multiset{x_i,x_{i+1}}$ or $q-1\in\multiset{x_i,x_{i+1}}$, $\bby$ is obtained from $\bfx$ by a single-deletion, which can be recovered since $\bfx\in\mathcal{C}_2(n;\bm{s},\bm{t})$.

 If $0,q-1\notin\multiset{x_i,x_{i+1}}$, we can determine the multiset $\multiset{x_i,x_{i+1}}$ and thus know whether $x_i+x_{i+1}<q$ or not. 
 We have two cases:
\begin{enumerate}
\item If $x_i+x_{i+1}<q$, the following algorithm can be used to recover $\bfx$. Scan the symbols from the beginning of $\bby$ to its end. If the symbol $c$ is encountered, conduct the following steps.
\begin{enumerate}[\textbf{Step} 1]
  \item If $x_ix_{i+1}=aa$ for some $a$, replace this $c$ with $aa$. Denote the resulting sequence by $\bfx^{\prime}$ and go to Step 4. 
  
  If $x_i\ne x_{i+1}$, we must have $x_ix_{i+1}\in\mathset{ab,ba}$ for some $a\ne b$. Go to Step 2.
  \item Replace this $c$ with $ab$ and denote the resulting sequence by $\bfx^{\prime}$. If $\Inv{\bfx^{\prime}}\equiv t_2\pmod{2}$, go to Step 4. Otherwise, keep this $c$ unchanged and go to Step 3.
  \item Replace this $c$ with $ba$. Denote the resulting sequence by $\bfx^{\prime}$ and go to Step 4. Otherwise, keep this $c$ unchanged and go to Step 5.
  \item If $\Syn{\bfx^{\prime}}\equiv d_1\pmod{qn}$, let $\bfx=\bfx^{\prime}$ and output $\bfx$. Otherwise, keep this $c$ unchanged and go to Step 5.
  \item Move to the next $c$ and go to Step 1.
\end{enumerate}
 Since $\bby$ is obtained from $\bfx$ by replacing an $x_ix_{i+1}$ with $c$, this $\bfx^{\prime}$ does exist. On the other hand, \Lref{lem_smallvalue} ensures that such $\bfx^{\prime}$ is unique.
  \item If $x_i+x_{i+1}\ge q$, we can recover $\bfx$ by the following procedure. First, since $P\parenv{\bfx}\in\cC_3(n;d_1)$, we can determine the error position $i$ from $P\parenv{\bby}$ by the algorithm given in the proof of \Lref{lem_largevalue}. If $x_ix_{i+1}=aa$ for some $a$, then replace $y_i$ ($=q-1$) with $aa$ and output the resulting sequence. If $x_ix_{i+1}\in\mathset{ab,ba}$ for some $a\ne b$, we can know whether $x_ix_{i+1}=ab$ or $x_ix_{i+1}=ba$ by $\Inv{\bfx}\pmod{2}$. Once $x_ix_{i+1}$ is determined, replace $y_i$ with $x_ix_{i+1}$ and output the resulting sequence. \Lref{lem_largevalue} ensures that the sequence $\bfx$ can be uniquely recovered.
\end{enumerate}
\end{IEEEproof}

By the pigeonhole principle, there are some $\bm{s}$, $\bm{t}$ and $\bm{d}$ such that 
\begin{equation}\label{eq_sizeofcodes}
\abs{\mathcal{C}(n;\bm{s},\bm{t},\bm{u})}\ge\frac{q^n}{4^{q-1}\cdot n\cdot 2\cdot\parenv{qn}\cdot \parenv{2n-3}}.
\end{equation}
This lower bound means that the redundancy of $\mathcal{C}(n;\bm{s},\bm{t},\bm{u})$
is at most $3\log_q(n)+O(1)$ for some choice of $\bm{s}$, $\bm{t}$ and $\bm{d}$. If measured in binary bits, this redundancy is at most $3\log_2(n)+O(1)$.
Recall that in \cite[Corollary 17]{Yaakobi2020isit}, the authors gave a $q$-ary single-deletion single-substitution correcting code of redundancy at most $10\log_2(n)+O(1)$. So the code   $\mathcal{C}(n;\bm{s},\bm{t},\bm{d})$ performs better than the existing one in \cite{Yaakobi2020isit}. One may ask if $3\log_q(n)+O(1)$ is the best redundancy that can be achieved. Based on $\cC\parenv{n;\bm{s},\bm{t},\bm{d}}$ and new ideas, we will show in next subsection that the redundancy can be further reduced to at most $\log_q(n)+O(\log_q\log_q(n))$.

\subsection{An improved construction}\label{sec_improved}
In this subsection, we use \Tref{thm_primarycode} together with ideas from \cite{Gabrys2022IT} and provide a code with redundancy $\log_q(n)+O(\log_q\log_q(n))$. We first outline the basic idea.

We begin with constructing a code with redundancy $\log_q(n)+O(1)$. This code has the property that when receiving a sequence which is a corrupted version of a codeword $\bfx$, it is possible to locate a \underline{window} of length $L=\Theta\left(\log_q^2(n)\right)$ that contains the erroneous position. 
That is to say, we only need to correct the absorption error within a shorter substring of $\bfx$. 
To this end, we should partition $\bfx$ into consecutive disjoint intervals of length $2L+1$ and then apply \Tref{thm_primarycode} to each of these intervals. As we will show next, this will only increase the redundancy by $O(\log_q\log_q(n))$ and so the overall redundancy of the resulted code is $\log_q(n)+O(\log_q\log_q(n))$. The details will be clear from the subsequent analysis.

For each $\bfx\in\Sigma_q^{n}$, which ends with $0011$, we can segment $\bfx$ and get a string $\bbz^{\bfx}=\bbz_1^{\bfx}\cdots \bbz_{l_{\bfx}}^{\bfx}$, where $1\le l_{\bfx}\le n/4$, and each substring $\bbz_i^{\bfx}$ ends with $0011$, and $0011$ appears exactly once in $\bbz_i^{\bfx}$. For example, let $q=3$ and $\bfx=00111230320011$. Then $l_{\bfx}=2$ and $\bbz_1^{\bfx}=0011$, $\bbz_2^{\bfx}=1230320011$.

Let $\delta=c_1+c_2\ceilenv{\log_q(n)}$, where constants $c_1$ and $c_2$ are both multiples of $4$ and satisfy
\begin{equation*}
  \left(\frac{q^4}{q^4-1}\right)^{\frac{c_1}{4}-1}\ge\frac{q}{q-1},\text{ and } \left(\frac{q^4}{q^4-1}\right)^{\frac{c_2}{4}}\ge q.
\end{equation*}
Since $\frac{q^4}{q^4-1}>1$, the desired constants $c_1$ and $c_2$ do exist. For example, if $q=3$, the smallest $c_1$ is $136$, while the smallest $c_2$ is $356$.
\begin{lemma}\label{lem_shortmarker}
  Suppose that $X$ is chosen uniformly at random from $\Sigma_q^n$. Then
  \begin{equation*}
    \Pr\left(\left|\bbz_i^X\right|\le\delta,i=1,\ldots,l_x\right)\ge\frac{1}{q}.
  \end{equation*}
\end{lemma}
\begin{IEEEproof}
The probability that a fixed length-$4$ substring of $X$ equals $0011$ is $\frac{1}{q^4}$. Then for any $i$, the probability that $\left|\bbz_i^X\right|>\delta$ is at most
  \begin{equation*}
    \left(\frac{q^4-1}{q^4}\right)^{\frac{\delta-4}{4}}\le\frac{q-1}{qn},
  \end{equation*}
  where the inequality follows from the choices of $c_1$ and $c_2$. Now the conclusion follows from the union bound.
\end{IEEEproof}

Let $\mathcal{R}_{q,n}$ be the set of all strings $\bfx\in\Sigma_q^{n}$ which ends with $0011$ and satisfies the condition that $\left|\bbz_i^X\right|\le\delta$ for all $i=1,\ldots,l_{\bfx}$.  Then \Lref{lem_shortmarker} implies $\left|\mathcal{R}_{q,n}\right|\ge q^{n-5}$. Next, we briefly explain how to construct $\cR_{q,n}$. Let $$
\cZ=\mathset{\bbz\in\Sigma_q^{\le\delta}~:\bbz\text{ ends with }0011\text{ and }0011\text{ does not appear elsewhere in }\bbz }.
$$
Since $\delta=c_1+c_2\ceilenv{\log_q(n)}$, the size of $\Sigma_q^{\le\delta}$ is bounded above by $O(n^{c_2})$\footnote{More accurately, the capacity of the set of strings of length $n$ that do not contain $0011$, which can be calculated using constrained systems techniques, is $\log_q (1.839)$ which is roughly $0.87$ in the binary case.}. This implies that $\cZ$ can be constructed by brute force searching. We can construct $\cR_{q,n}$ by concatenating sequences in $\cZ$. This process can be somewhat involved, but this is a one-time pre-processing task. When $(c_1-4)\log_q(e)/(4q^4)\ge 5$ and $c_2\log_q(e)/(4q^4)\ge 1$, we present an algorithm for encoding(and decoding) an arbitrary sequence of length $n$ into a sequence in $\cR_{q,n+5}$ (see Appendix \ref{sec_algorithms}).

\begin{observation}\label{obs_marker}
  Let $\bfx\in\Sigma_q^{n}$ be a string, ending with $0011$. If the last $0011$ is destroyed due to an absorption error, it is easy to detect and correct that error. If the absorption error does not change the last $0011$ and the received sequence is $\bby$, we have $|\bby|=|\bfx|-1$ and $l_{\bby}-l_{\bfx}\in\{0,-1,1\}$ where $l_{\bfx},l_{\bby}$ denote the number of substrings that end with $0011$ in $\bfx,\bby$, respectively. 
\end{observation}

For any $n\in\N$ and any $\bfx\in\Sigma_q^n$, define
\begin{equation*}
	\begin{array}{l}
	f(\bfx)=\mathop{\sum}\limits_{j=1}^{l_{\bfx}}j\left|\bbz_j^{\bfx}\right|\pmod{2n},\\
		g(\bfx)=l_{\bfx}\pmod{3}.
	\end{array}
\end{equation*}
If $\bby$ is obtained from $\bfx$ by an absorption, the function $g(\bfx)$ can help us to determine the exact value of $l_{\bby}-l_{\bfx}$. For a given $\bm{r}=\parenv{r_1,r_2}\in \mathbb{Z}_{2n}\times \mathbb{Z}_3$, we define the code $\mathcal{D}_1(n;\bm{r})\subseteq\Sigma_q^n$ as
\begin{equation*}
  \mathcal{D}_1(n;\bm{r})=\left\{\bfx\in\mathcal{R}_{q,n}:\;f(\bfx)=r_1,g(\bfx)=r_2\right\}.
\end{equation*}
With suitable parameters, this code has redundancy at most $\log_q(n)+O(1)$. 

\begin{theorem}\label{thm_window}
  Let $\bfx\in\mathcal{D}_1(n;\bbr)$ be a sequence and let $\bby$ be the sequence obtained from $\bfx$ after a single absorption. Then there is a constant $c_3$, which is a function of $c_1$ and $c_2$, such that a window $W\subseteq[1,n-1]$ of size $c_3\log_q^2(n)$ that contains the position where the absorption error has occurred in $\bby$, can be detected. Furthermore, the  window can be found in $O(n)$ time.
\end{theorem}
The proof of \Tref{thm_window} is similar to the proof of \cite[Theorem 4]{Gabrys2022IT} and is deferred to Appendix \ref{appendix1}.

Let $L=c_3\log_q^2(n)$. For simplicity, we assume $(2L+1)\mid n$  and let $t=n/(2L+1)$. All the following arguments can be generalized to the case $(2L+1)\nmid n$ in a straightforward way (see \Remarkref{rmk_notdivide} below). 
We partition $\{1,\ldots,n\}$ into consecutive disjoint intervals $I_1^{(1)},\ldots,I_t^{(1)}$ of length $2L+1$. In other words,
\begin{equation}\label{eq_intervals}
I_i^{(1)}=\sparenv{1+(i-1)(2L+1),i(2L+1)}
\end{equation}
for all $1\le i\le t$.
Furthermore, we define a family of shifted intervals $I_1^{(2)},\ldots,I_{t-1}^{(2)}$, where $I_i^{(2)}=I_i^{(1)}+L$.\footnote{For a set $A$ of integers and an integer $m$, we define $A+m=\mathset{a+m~:~a\in A}$.}
For given $\bfx\in\Sigma_q^n$, let $\bfx^{(1,i)}=\bfx_{I_i^{(1)}}$ and $\bfx^{(2,i)}=\bfx_{I_i^{(2)}}$. In other words, $\bfx^{(1,i)}$ is the substring corresponding to $I_i^{(1)}$ and $\bfx^{(2,i)}$ is the substring corresponding to $I_i^{(2)}$.

For a given sequence $\bbz\in\Sigma_q^{2L+1}$, we define
\begin{align*}
  \hat{f}(\bbz)=\left(N_a(\bbz)\right)_{a\in[0,q-2]}\times\parenv{\Syn{\alpha(\bbz)},\Inv{\bbz},\Syn{\bbz},\Syn{P\parenv{\bbz}}}.
\end{align*}
The values of $\hat{f}(\bbz)$ are taken from $\mathbb{Z}_4^{q-1}\times\mathbb{Z}_{2L+1}\times\mathbb{Z}_2\times\mathbb{Z}_{q(2L+1)}\times\mathbb{Z}_{4L-1}$.
With the function $\hat{f}(\cdot)$ in hand, we define the functions:
\begin{equation*}
\begin{array}{l}
\hat{g}_1(\bfx)=\mathop{\sum}\limits_{i=1}^{t}\hat{f}\left(\bfx^{(1,i)}\right), \\
  \hat{g}_2(\bfx)=\mathop{\sum}\limits_{i=1}^{t-1}\hat{f}\left(\bfx^{(2,i)}\right),
\end{array}
\end{equation*}
where the sums are performed position-wise over $\mathbb{Z}_4^{q-1}\times\mathbb{Z}_{2L+1}\times\mathbb{Z}_2\times\mathbb{Z}_{q(2L+1)}\times\mathbb{Z}_{4L-1}$. 
Now we can give the desired code. For given $\bm{\alpha}$, $\bm{\beta}$ $\in$ $\mathbb{Z}_4^{q-1}\times\mathbb{Z}_{2L+1}\times\mathbb{Z}_2\times\mathbb{Z}_{q(2L+1)}\times\mathbb{Z}_{4L-1}$ and $\bm{r}=\parenv{r_1,r_2}\in \mathbb{Z}_{2n}\times \mathbb{Z}_3$, let
\begin{equation*}
  \mathcal{D}(n;\bm{r},\bm{\alpha},\bm{\beta})=\mathcal{D}_1(n;\bm{r})\bigcap\mathset{\bfx\in\mathcal{R}_{q,n}:\;\hat{g}_1(\bfx)=\bm{\alpha},\hat{g}_2(\bfx)=\bm{\beta}}.
\end{equation*}
Similar to \Cref{eq_sizeofcodes}, there exists a choice of $\bm{r}$, $\bm{\alpha}$ and $\bm{\beta}$, such that
$$
\abs{\mathcal{D}(n;\bm{r},\bm{\alpha},\bm{\beta})}\ge\frac{\abs{\cR_{q,n}}}{2n\cdot 3\cdot\sparenv{4^{q-1}\cdot\parenv{2L+1}\cdot 2\cdot \parenv{q(2L+1)}\cdot\parenv{4L-1}}^2}.
$$
Therefore, the redundancy of $\mathcal{D}(n;\bm{r},\bm{\alpha},\bm{\beta})$ is at most
$\log_q(n)+12\log_q\log_q(n)+O(1)$ (recall that we require that $c_1$, $c_2$ and $c_3$ are constants and $n$ is large compared to these constants).

\begin{theorem}\label{thm_nonbinarycode}
  The code $\mathcal{D}(n;\bm{r},\bm{\alpha},\bm{\beta})$ can correct a single absorption error.
\end{theorem}
\begin{IEEEproof}
  Let $\bfx$ be the transmitted codeword and $\bby$ be the received sequence. The proof of \Tref{thm_window} gives a method to locate the error position within a window $W=\sparenv{i_1,i_1+L-1}\subseteq[n-1]$. By the constructions of $I_i^{(1)}$'s and $I_i^{(2)}$'s, there exists some $i$ such that $W$ is contained in $I_i^{(1)}$ or $I_i^{(2)}$. The value of $i$ can be determined in the following way (recall \Cref{eq_intervals} for the definitions of $I_i^{(1)}$'s and $I_i^{(2)}$'s).
  \begin{enumerate}[\textbf{Step }1]
      \item Find the largest $k\ge0$ such that $k(2L+1)<i_1$ and $W\subseteq\sparenv{k(2L+1)+1,k(2L+1)+2L}$. Then $i=k+1$. If such a $k$ does not exist, go to Step 2.
      \item Find the largest $k\ge0$ such that $k(2L+1)+L<i_1$ and $W\subseteq\sparenv{k(2L+1)+L+1,k(2L+1)+3L}$. Then $i=k+1$.
  \end{enumerate}
  Since any window of length $L$ must be contained in some $I_i^{(1)}$ or $I_i^{(2)}$, the above two steps can successfully find such an $i$. Now we can recover $\bfx$ from $\bby$ by the following procedure.
  \begin{enumerate}[\textbf{Case }(1)]
    \item The value of $i$ is found in Step 1. In this case, we have $x_j=y_j$ for all $j\le(i-1)(2L+1)$ and $x_j=y_{j-1}$ for all $j>i(2L+1)$. In other words, we can recover $\bfx^{(1,j)}$ for all $j\ne i$ directly. Therefore, we can compute $\hat{f}(\bfx^{(1,j)})$ for all $j\ne i$. Then comparing $\bm{\alpha}$ and $\mathop{\sum}\limits_{j\ne i}\hat{f}\parenv{\bfx^{(1,j)}}$, we can know $\hat{f}\parenv{\bfx^{(1,i)}}$. Let $\bby^{(i)}=\bby_{\sparenv{(i-1)(2L+1)+1,i(2L+1)-1}}$. Then $\bby^{(i)}$ is the corrupted version of $\bfx^{(1,i)}$. \Cref{thm_primarycode} ensures that we can recover $\bfx^{(1,i)}$ from $\bby^{(i)}$ with the help of $\hat{f}\parenv{\bfx^{(1,i)}}$. Now the transmitted sequence $\bfx$ is recovered.
    \item The value of $i$ is found in Step 2. In this case, we have $x_j=y_j$ for all $j\le(i-1)(2L+1)+L$ and $x_j=y_{j-1}$ for all $j>i(2L+1)+L$. In other words, we can recover $x_j$ for all $j\notin\sparenv{(i-1)(2L+1)+L+1,i(2L+1)+L}$ directly. Therefore, we can compute $\hat{f}(\bfx^{(2,j)})$ for all $j\ne i$. Then comparing $\bm{\beta}$ and $\mathop{\sum}\limits_{j\ne i}\hat{f}\parenv{\bfx^{(2,j)}}$, we can know $\hat{f}\parenv{\bfx^{(2,i)}}$. Let $\bby^{(i)}=\bby_{\sparenv{(i-1)(2L+1)+L+1,i(2L+1)+L-1}}$. Then $\bby^{(i)}$ is the corrupted version of $\bfx^{(2,i)}$. \Cref{thm_primarycode} ensures that we can recover $\bfx^{(2,i)}$ from $\bby^{(i)}$ with the help of $\hat{f}\parenv{\bfx^{(2,i)}}$. Now the transmitted sequence $\bfx$ is recovered.
  \end{enumerate}
\end{IEEEproof}
\begin{remark}\label{rmk_notdivide}
    If $2L+1\nmid n$, let $t=\floorenv{n/(2L+1)}$ and $L^{\prime}=n-t(2L+1)$. Then $0<L^{\prime}\le 2L$. The $2t-1$ intervals $I_i^{(1)}$ ($1\le i\le t$) and $I_i^{(2)}$ ($1\le i\le t-1$) are defined as above. There are two cases.
    \begin{itemize}
        \item When $L^{\prime}\le L$, let $I_t^{(2)}=\sparenv{(t-1)(2L+1)+L+1,n}$. Then $L+2\le\abs{I_t^{(2)}}\le 2L+1$. So we define $\hat{g}_1(\bfx)$ as above and $\hat{g}_2(\bfx)=\mathop{\sum}\limits_{i=1}^{t}\hat{f}\parenv{\bfx^{(2,i)}}$.
        \item When $L<L^{\prime}\le 2L$, let $I_{t+1}^{(1)}=\sparenv{t(2L+1)+1,n}$ and $I_t^{(2)}=\sparenv{(t-1)(2L+1)+L+1,t(2L+1)+L}$. Then $L<\abs{I_{t+1}^{(1)}}\le 2L$ and $\abs{I_t^{(2)}}=2L+1$. So we define $\hat{g}_1(\bfx)=\mathop{\sum}\limits_{i=1}^{t+1}\hat{f}\parenv{\bfx^{(1,i)}}$ and $\hat{g}_2(\bfx)=\mathop{\sum}\limits_{i=1}^{t}\hat{f}\parenv{\bfx^{(2,i)}}$.
    \end{itemize}
\end{remark}

\subsection{Codes correcting multiple errors}\label{sec_multiple}
In this subsection, we study codes that can correct multiple absorption errors. Recall that the alphabet size $q$ is at least $3$, unless otherwise stated.
We first claim that $t$-absorption is a special case of $t$-deletion-$t$-substitution and give two known results. After that, we explain the difference between $t$-absorption and  $t$-deletion-$t$-substitution, which justifies our searching for better codes for our setting. Our construction is based on the single-absorption correcting code given in \Cref{thm_nonbinarycode} and the syndrome compression technique with precoding developed recently \cite{Song2022IT}.

In \Obsref{obs_general}, we have shown that a single-absorption error corresponds to a single-deletion together with at most a single-substitution. By \Dref{dfn_absorption}, it is not difficult to see that this conclusion holds for multiple absorptions as well. In other words, a $t$-absorption error is the combination of $t$ deletions and \emph{at most} $t$ substitutions. To see that, it suffices to notice that the absorption error $\parenv{\mathop{\oplus}\limits_{j=i_l}^{i_l+s_l}x_j}$ can be interpreted as firstly deleting $s_l$ symbols $x_j$ ($i_1\le j<i_l+s_l$) and then substituting $x_{i_l+s_l}$ by $\parenv{\mathop{\oplus}\limits_{j=i_l}^{i_l+s_l}x_j}$. If $\parenv{\mathop{\oplus}\limits_{j=i_l}^{i_l+s_l}x_j}\ne x_{i_l+s_l}$, the second step is a substitution error. In other words, the absorption error $\parenv{\mathop{\oplus}\limits_{j=i_l}^{i_l+s_l}x_j}$ of $s_l+1$ consecutive symbols can be interpreted as $s_l$ deletions and \emph{at most} one substitution.

Therefore, a $t$-deletion-$t$-substitution correcting code is naturally a $t$-absorption correcting code.
We first introduce two classes of $t$-deletion-$t$-substitution correcting codes given in the literature. They will be used as a bulding block in our construction of $t$-absorption correcting codes.

By carefully checking the proof of \cite[Lemma 9]{Song2022IT}, we draw the following conclusion.
\begin{lemma}\label{lem_DScode1}
  Suppose that $q\ge3$ and $t$ are fixed positive integers. There exists a $q$-ary systematic\footnote{In the proof of \cite[Lemma 9]{Song2022IT}, an systematic encoder was defined.} $t$-deletion $t$-substitution correcting code $\cE_q\subseteq\Sigma_q^N$ whose redundancy is at most $\frac{22t}{\log_q(2)}\log_q(N)+o\parenv{\log_q(N)}$. The encoding and decoding complexities\footnote{These two complexities follow from the construction of $\cE_q$ and \cite[Theorem 1]{Song2022IT}.} are $O\parenv{N^{6t+1}}$ and $O\parenv{N^{3t+1}}$, respectively.
\end{lemma}

When $q$ is a prime power\footnote{When constructing the code in \cite[Theorem 3]{Song2022IT}, the authors used a BCH code over the finite field $\mathbb{F}_q$. This is the reason why we require that $q$ is a prime power.}, the authors of \cite{Song2022IT} obtained a better result. 
\begin{lemma}\cite[Theorem 3]{Song2022IT}\label{lem_DScode2}
Let $q\ge 3$ be a prime power. There exists a $q$-ary systematic $t$-deletion $t$-substitution correcting code $\cE_q\subseteq\Sigma_q^N$ with redundancy at most $\parenv{8t-1-\floorenv{\frac{2t-1}{q}}}\log_q(N)+o\parenv{\log_q(N)}$. The encoding and decoding complexities are $O\parenv{N^{4t+1}}$ and $O\parenv{N^{2t+1}}$, respectively.
\end{lemma}

Furthermore, the codes $\cE_q$ in \cite[Lemma 9]{Song2022IT} and \cite[Theorem 3]{Song2022IT} can be expressed as
\begin{equation}\label{eq_DScode}
\begin{aligned}
  \cE_q=\bracenv{\parenv{\bbu,\Red{q,n}{\bbu}}~:~\bbu\in\Sigma_q^n}
\end{aligned}
\end{equation}
where $\bbu$ is the information sequence and $\Red{q,n}{\bbu}$ is the sequence of redundancy symbols. Note that $N=n+\abs{\Red{q,n}{\bbu}}$. Let
\begin{equation*}
R_{q,n}=
\begin{cases}
  \parenv{8t-1-\floorenv{\frac{2t-1}{q}}}\log_q(N)+o\parenv{\log_q(N)} & \mbox{if }q\text{ is a prime power}, \\
  \frac{22t}{\log_q(2)}\log_q(N)+o\parenv{\log_q(N)}, & \mbox{if }q\text{ is arbitrary}.
\end{cases}
\end{equation*}
Since $\frac{n}{N}\ge\frac{1}{2}$ when $n$ is sufficiently large, we have
\begin{equation}\label{eq_valueofRq}
R_{q,n}=
\begin{cases}
  \parenv{8t-1-\floorenv{\frac{2t-1}{q}}}\log_q(n)+o\parenv{\log_q(n)} & \mbox{if }q\text{ is a prime power}, \\
  \frac{22t}{\log_q(2)}\log_q(n)+o\parenv{\log_q(n)}, & \mbox{otherwise}.
\end{cases}
\end{equation}
In the following, whenever $R_{q,n}$ is mentioned, we always refer to \Cref{eq_valueofRq}.

From \Lref{lem_DScode1} and \Lref{lem_DScode2} we can see that $\Red{q,n}{\bbu}\in\Sigma_q^{R_{q,n}}$. For our purpose, we can also view $\red{q,n}$ as a function $\red{q,n}:\Sigma_q^n\rightarrow\sparenv{0,q^{R_{q,n}}-1}$. Let $\cB_t^{DS}\parenv{\bbu}$ be the $t$-deletion-$t$-substitution ball centered at $\bbu$, i.e.,
\begin{equation}\label{eq_DSball}
\cB_t^{DS}\parenv{\bbu}=
\left\{\bbz\in\Sigma_q^{n-t}~:~
  \begin{array}{l}
  \bbz \text{ is obtained from }\bbu\text{ by }t\text{ deletions}\\
  \text{and at most }t\text{ substitutions}
  \end{array}
  \right\}.
\end{equation}
Then \Lref{lem_DScode1}, \Lref{lem_DScode2} and \Cref{eq_DScode} imply the following corollary. 
\begin{corollary}
  \label{cor_simple1}
  If $\cB_t^{DS}\parenv{\bbu}\cap \cB_t^{DS}\parenv{\bbu^{\prime}}\ne\emptyset$ and $\bbu\ne\bbu^{\prime}$, then $\Red{q,n}{\bbu}\ne\Red{q,n}{\bbu^{\prime}}$.
\end{corollary}

As discussed above, \Lref{lem_DScode1} and \Lref{lem_DScode2} provide us with two class of $t$-absorption correcting codes with low redundancy. 
However, the two kinds of error models differ in the following two aspects.
\begin{itemize}
    \item In the $t$-deletion-$t$-substitution setup, the error positions are assumed to be arbitrary. But for the $t$-absorption channel, the deletion-positions and the substitution-positions are ``close". For example, the absorption error $\parenv{\mathop{\oplus}\limits_{j=i_l}^{i_l+s_l}x_j}$ leads to deletions in positions $j$ ($i_1\le j<i_l+s_l$) and a (possible) substitution in position $i_l+s_l$. Therefore, the deletions and substitution are constrained to within a window of length $s_l+1$.
    \item In the $t$-deletion-$t$-substitution setup, a symbol $a\in\Sigma_q$ can be substituted by an arbitrary symbol $b\in\Sigma_q\setminus\{a\}$. However, for absorption channels, a symbol $a\in\Sigma_q$ can only be substituted by some $b>a$ and $b\in\Sigma_q$.
\end{itemize}

Therefore, it is reasonable to deem that there are better codes for absorption channels, which is the main goal of this subsection.
In the rest of this subsection, we will apply the syndrome compression technique with precoding to show that for our setting, there are codes with even lower redundancy. The syndrome compression technique was first established in \cite{Sima2019isit,Sima2021it} for designing $t$-deletion correcting codes, and then was further developed in \cite{Sima2020isit} to a general method for obtaining low-redundancy error correcting codes. More recently, \cite{Song2022IT} further improved the syndrome compression technique by applying a precoding process.

To describe the syndrome compression technique, we need to introduce some notations. Let $\cB\parenv{\bbu}$ be a general error ball centered at the sequence $\bbu\in\Sigma_q^n$. The definition of such error balls is determined by the specific problem under consideration. For example, if we are studying $t$-deletion-$t$-substitution error correcting codes, then the error ball $\cB\parenv{\bbu}$ is defined as \Cref{eq_DSball}. Consider some fixed error and its corresponding error ball $\cB(\bbu)$. For a given code $\cE\subseteq\Sigma_q^n$ and $\bbu\in\cE$, we define
$$
\cN_{\cE}\parenv{\bbu}=\mathset{\bbu^{\prime}\in\cE~:~\bbu^{\prime}\ne\bbu\text{ and }\cB\parenv{\bbu^{\prime}}\cap\cB\parenv{\bbu}\ne\emptyset}.
$$

The following lemma, which is a variant of \cite[Lemma 1]{Sima2019isit} and \cite[Lemma 3]{Song2022IT}, is key to our purpose. We include its proof here because the proof reveals how the syndrome compression technique works.
\begin{lemma}\label{lem_compression}
Let $\cE\subseteq\Sigma_q^n$ be a code and $N>\max\bracenv{\abs{\cN_{\cE}\parenv{\bbu}}~:~\bbu\in\cE}$.
Suppose that the function $f:\Sigma_q^n\rightarrow\sparenv{0,q^{R(n)}-1}$ (where $R(n)$ is a function of $n$ and $R(n)\ge2$) satisfies the following property:
\begin{enumerate}[$(\textup{\textbf{P}} 1)$]
    \item if $\bbu\in\Sigma_q^n$ and $\bbu^{\prime}\in\cN_{\Sigma_q^n}\parenv{\bbu}$, then $f\parenv{\bbu}\ne f\parenv{\bbu^{\prime}}$.
\end{enumerate}
Then there exists a function $\bar{f}:\cE\rightarrow\sparenv{0,q^{2\log_q(N)+O\parenv{\frac{R(n)}{\log_q\parenv{R(n)}}}}-1}$ such that $\bar{f}\parenv{\bbu}\ne \bar{f}\parenv{\bbu^{\prime}}$ for any $\bbu\in\cE$ and $\bbu^{\prime}\in\cN_{\cE}\parenv{\bbu}$.
\end{lemma}
\begin{IEEEproof}
 For any $\bbu\in\cE$ and $\bbu^{\prime}\in\cN_{\cE}\parenv{\bbu}$, we have $1\le\abs{f\parenv{\bbu}-f\parenv{\bbu^{\prime}}}<q^{R(n)}$ due to (\textbf{P}1). For any $\bbu\in\cE$, let
  $$
  D\parenv{\bbu}=\mathset{p~:~p\text{ is a positive divisor of }\abs{f\parenv{\bbu}-f\parenv{\bbu^{\prime}}}\text{ for some }\bbu^{\prime}\in\cN_{\cE}\parenv{\bbu}}.
  $$
  By \cite[Lemma 3]{Sima2020isit}, the number of positive divisors of $\abs{f\parenv{\bbu}-f\parenv{\bbu^{\prime}}}$ is upper bounded by
  $$
  q^{O\parenv{\frac{R(n)}{\log_q\parenv{R(n)}}}},
  $$
  for each $\bbu^{\prime}\in\cN_{\cE}\parenv{\bbu}$. So we have
  $$
  \abs{D\parenv{\bbu}}\le\abs{\cN_{\cE}\parenv{\bbu}}q^{O\parenv{\frac{R(n)}{\log_q\parenv{R(n)}}}}< Nq^{O\parenv{\frac{R(n)}{\log_q\parenv{R(n)}}}}=q^{\log_q(N)+O\parenv{\frac{R(n)}{\log_q\parenv{R(n)}}}}.
  $$
  This implies that there is an integer $P\parenv{\bbu}\in\sparenv{1,q^{\log_q(N)+O\parenv{\frac{R(n)}{\log_q\parenv{R(n)}}}}}$ such that $f\parenv{\bbu}\not\equiv f\parenv{\bbu^{\prime}}\pmod{P\parenv{\bbu}}$ for all $\bbu^{\prime}\in\cN_{\cE}(\bbu)$. Now for each $\bbu\in\cE$, we define
  \begin{equation*}
    \bar{f}\parenv{\bbu}=\parenv{\Expan{q}{f\parenv{\bbu}\pmod{P\parenv{\bbu}}},\Expan{q}{P\parenv{\bbu}}},
  \end{equation*}
  where $\Expan{q}{m}$ is the $q$-ary expansion of the integer $m$. Clearly, $\bar{f}\parenv{\bbu}$ is a $q$-ary vector of length $2\log_q(N)+O\parenv{\frac{R(n)}{\log_q\parenv{R(n)}}}$ and thus we can view $\bar{f}$ as a function $\bar{f}:\cE\rightarrow\sparenv{0,q^{2\log_q(N)+O\parenv{\frac{R(n)}{\log_q\parenv{R(n)}}}}-1}$. By construction, it holds that $\bar{f}\parenv{\bbu}\ne \bar{f}\parenv{\bbu^{\prime}}$ for any $\bbu\in\cE$ and $\bbu^{\prime}\in\cN_{\cE}\parenv{\bbu}$.
\end{IEEEproof}
\begin{remark}\label{rmk_barf}
  In most cases,  the number $N$ is a polynomial in $n$. So if it holds that $O\parenv{\frac{R(n)}{\log_q\parenv{R(n)}}}$ $=$ $O\parenv{\log_q(n)}$, the function $\bar{f}$ can be computed in polynomial time.
\end{remark}

Before moving on, we explain how \Lref{lem_compression} helps to compress the code redundancy. We follow the notations in \Lref{lem_compression}. For a given $a_1\in\sparenv{0,q^{R(n)}-1}$, the function $f$ can be used to define a code 
$$
\cE^{\prime}\parenv{a_1}=\mathset{\bbu\in\Sigma_q^n~:~f\parenv{\bbu}=a_1},
$$
where there exists some $a_1$ such that the redundancy of $\cE^{\prime}\parenv{a_1}$ is at most $R(n)$. If the conditions in \Lref{lem_compression} are satisfied, then the function $\bar{f}$ can be used to define another code
$$
\cE^{\prime\prime}\parenv{a_2}=\mathset{\bbu\in\cE~:~\bar{f}\parenv{\bbu}=a_2}
$$
where $a_2\in\sparenv{0,q^{2\log_q(N)+O\parenv{\frac{R(n)}{\log_q\parenv{R(n)}}}}-1}$, and there exists some $a_2$ such that the redundancy of $\cE^{\prime\prime}\parenv{a_2}$ is at most $r\parenv{\cE}+2\log_q(N)+O\parenv{\frac{R(n)}{\log_q\parenv{R(n)}}}$, where $r\parenv{\cE}$ is the redundancy of $\cE$.  If $r\parenv{\cE}+2\log_q(N)+O\parenv{\frac{R(n)}{\log_q\parenv{R(n)}}}$ is much smaller than $R(n)$, then the code redundancy is successfully compressed. If $\cE=\Sigma_q^n$, we obtain the original syndrome compression technique in \cite{Sima2020isit}. If $\cE$ is chosen to be a proper subset of $\Sigma_q^n$, then we obtain the syndrome compression technique with precoding in \cite{Song2022IT}.

Now we are ready to derive the main result of this subsection, that is, $t$-absorption correcting codes ($t\ge2$). In this case, the error ball $\cB\parenv{\bbu}$ is defined to be the $t$-absorption ball (see \Cref{eq_absorptionball}), i.e.,
$$
\cB\parenv{\bbu}=\cB_t^{ab}\parenv{\bbu}=\mathset{\bbz\in\Sigma_q^{n-t}~:~\bbz\text{ is obtained from }\bbu\text{ by }t\text{ absorption errors}}.
$$
We choose $\cE$ to be the code $\mathcal{D}\parenv{n;\bm{r},\bm{\alpha},\bm{\beta}}$ in \Cref{thm_nonbinarycode}, and $f$ to be the function $\red{q,n}$ (see \Cref{eq_DScode}). From Corollary \ref{cor_simple1}, $f$ satisfies the property (\textbf{P}1) in \Lref{lem_compression} with $R(n)=R_{q,n}$, which is defined as in \Cref{eq_valueofRq}. So we have $O\parenv{\frac{R(n)}{\log_q\parenv{R(n)}}}$ $=$ $o\parenv{\log_q(n)}$.

Firstly, we need to estimate an upper bound of $\abs{\cN_{\cE}\parenv{\bbu}}$ for any $\bbu\in\cE$. For a given sequence $\bbz$, if we insert a symbol at the end of $\bbz$, or replace some $z_i$ with $ab$ such that $z_i=a\oplus b$, then we say we perform a \emph{splitting} operation on $\bbz$.
\begin{claim}\label{clm_confusionball}
Let $t\ge2$. Then for any $\bbu\in\cE$, we have $\abs{\cN_{\cE}\parenv{\bbu}}<q^{2t-2}n^{2t-1}$.
\end{claim}
\begin{IEEEproof}
  We should estimate the number of $\bbu^{\prime}\in\cE$ such that $\bbu^{\prime}\ne\bbu$ and $\cB_t\parenv{\bbu^{\prime}}\cap\cB_t\parenv{\bbu}\ne\emptyset$. Each such $\bbu^{\prime}$ can be obtained through the following steps.
  \begin{enumerate}[\textbf{Step} 1]
    \item Obtain a sequence $\bbu^{(1)}$ from $\bbu$ by sequentially performing $t$ absorptions, which has at most $n(n-1)\cdots(n-t+1)<n^t$ possibilities.
    \item For each $\bbu^{(1)}$, we perform a splitting operation on $\bbu^{(1)}$ to get a sequence $\bbz^{(1)}$. Then we perform a splitting operation on $\bbz^{(1)}$ to get a sequence $\bbz^{(2)}$. Repeat this process. after $t-1$ steps, we will get a sequence $\bbz^{(t-1)}$. For each $\bbu^{(1)}$, there are at most $q^{2t-2}(n-t+1)(n-t+2)\cdots(n-1)<q^{2t-2}n^{t-1}$ such $\bbz^{(t-1)}$ 's.
    \item For each $\bbz^{(t-1)}$, we perform a splitting operation on $\bbz^{(t-1)}$ to get a sequence $\bbu^{\prime}\in\cE$. Since $\cE$ is a single-absorption correcting code, there is at most one $\bbu^{\prime}$ for each $\bbz^{(t-1)}$.
  \end{enumerate}
  Overall, the number of $\bbu^{\prime}$ is strictly less than $q^{2t-2}n^{2t-1}$ and thus $\abs{\cN_{\cE}\parenv{\bbu}}<q^{2t-2}n^{2t-1}$.
\end{IEEEproof}

Now we choose $N=q^{2t-2}n^{2t-1}$. Then by \Lref{lem_compression}, we have a function $\bar{f}:\cE\rightarrow\sparenv{0,q^{\parenv{4t-2}\log_q(n)+o(\log_q(n))}-1}$
such that $\bar{f}\parenv{\bbu}\ne \bar{f}\parenv{\bbu^{\prime}}$ for any $\bbu\in\cE$ and $\bbu^{\prime}\in\cN_{\cE}\parenv{\bbu}$. Combining the above discussions, we obtain the main result of this subsection.
\begin{theorem}\label{thm_multipleerror}
  Let $q\ge 3$ and $t\ge 2$ be fixed integers. For given $\bm{\alpha}$, $\bm{\beta}$ $\in$ $\mathbb{Z}_4^{q-1}\times\mathbb{Z}_{2L}\times\mathbb{Z}_2\times\mathbb{Z}_{q(2L+1)}\times\mathbb{Z}_{4L-1}$, $\bm{r}=\parenv{r_1,r_2}\in \mathbb{Z}_{2n}\times \mathbb{Z}_3$ and $0\le a<q^{\parenv{4t-2}\log_q(n)+o(\log_q(n))}$, let
  $$
  \cE\parenv{n;\bm{r},\bm{\alpha},\bm{\beta},a}=\mathset{\bbc\in\cD\parenv{n;\bm{r},\bm{\alpha},\bm{\beta}}~:~\bar{f}\parenv{\bbc}=a}.
  $$
  Then $\cE\parenv{n;\bm{r},\bm{\alpha},\bm{\beta},a}$ is a $t$-absorption correcting code. Furthermore, there is a choice of $\bm{\alpha},\bm{\beta},\bm{r}$ and $a$, such that the redundancy of $\cE\parenv{n;\bm{r},\bm{\alpha},\bm{\beta},a}$ is at most
  $$
  \parenv{4t-1}\log_q(n)+o(\log_q(n)).
  $$
\end{theorem}

Let $\bbc\in\cE\parenv{n;\bm{r},\bm{\alpha},\bm{\beta},a}$ and $\hat{\bbc}\in\cB_t^{ab}\parenv{\bbc}$. By applying splitting operations on $\hat{\bbc}$, we can find at most $q^{2t-2}n^{t-1}$ sequences in $\cD\parenv{n;\bm{r},\bm{\alpha},\bm{\beta}}$ (see the proof of \Cref{thm_nonbinarycode} and Steps 2--3 in the proof of Claim \ref{clm_confusionball}). Among these sequences, there is a unique sequence $\widetilde{\bbc}$ such that $\bar{f}\parenv{\widetilde{\bbc}}=a$, and thus $\bbc=\widetilde{\bbc}$. Since finding the sequences $\cD\parenv{n;\bm{r},\bm{\alpha},\bm{\beta}}$ takes polynomial time, together with Remark \ref{rmk_barf} we obtain  that the function $\bar{f}$ can be computed in polynomial time. Therefore, we can recover $\bbc$ from $\hat{\bbc}$ in polynomial time.

\begin{remark}
    We do not know if there exists an efficient encoder that can encode an arbitrary sequence into $\mathcal{D}(n;\bm{r},\bm{\alpha},\bm{\beta})$ (or $\cE\parenv{n;\bm{r},\bm{\alpha},\bm{\beta},a}$). Based on the results in this section, we can provide two, polynomial-time encodable and decodable, codes $\cE_1$ and $\cE_2$, which can combat single-absorption and multiple-absorption errors, respectively. Details are deferred to Appendix \ref{variantcodes}. 
Recall that the code $\cD\parenv{n;\bm{r},\bm{\alpha},\bm{\beta}}$ is defined with four functions $f$, $g$, $\hat{g}_1$ and $\hat{g}_2$. For any two codewords $\bbc$ and $\bbc^{\prime}$ in $\cD\parenv{n;\bm{r},\bm{\alpha},\bm{\beta}}$, we have
$$
\parenv{f(\bbc),g(\bbc),\hat{g}_1(\bbc),\hat{g}_2(\bbc)}=\parenv{f(\bbc^{\prime}),g(\bbc^{\prime}),\hat{g}_1(\bbc^{\prime}),\hat{g}_2(\bbc^{\prime})}.
$$
However, for two codewords in $\cE_1$, the above equation is not necessarily true. The same phenomenon holds for $\cE_2$.
\end{remark}

\section{Optimality of the codes}\label{sec_optimality}
In this section, we always assume $q\ge2$.
Let $\mathcal{C}_{max}\subseteq\Sigma_q^n$ be a code of maximum size that can correct a single absorption error. Let $\cB_n$ denote the set of all $n$-length sequences over $\Sigma_q\setminus \{0\}$, i.e., the sequences that do not contain the symbol $0$. From \Obsref{obs_binary} and \Obsref{obs_general}, we know that the code $\mathcal{C}_{max}\setminus\mathcal{B}_n$ can correct a single deletion of zero. So $\left|\mathcal{C}_{max}\right|\le A_{q,n}+\left|\mathcal{B}_n\right|=A_{q,n}+(q-1)^n$,
where $A_{q,n}$ denotes the maximum size of a code in $\Sigma_q^n\setminus\mathcal{B}_n$ that can correct a single deletion of zero. In this section, we will prove an upper bound of $A_{q,n}$, which implies that the codes given in the last two sections are optimal or near optimal in terms of redundancy. To that end, we follow the method proposed in \cite{Kulkarni2013}, of which the authors proved a nonasymptotic upper bound of the size of a deletion correcting code (rather than zero-deletion correcting codes which we are interested in). The basic idea is to interpret our problem of upper bounding the size of codes as a linear programming problem. Inspired by \cite{Kulkarni2013}, several researchers further developed this method and obtained many important results (see, for example, \cite{Fazeli2015it,Cullina2016it}).

We need to introduce some terminologies first. A \textit{hypergraph} $\mathcal{H}$ is a tuple $\left(V,\mathcal{E}\right)$, where $V$ is a finite nonempty set and $\mathcal{E}$ is a collection of nonempty subsets of $V$. The set $V$ is the \textit{vertex set} of $\mathcal{H}$ and the elements in $V$ are called \textit{vertices}. The elements in $\mathcal{E}$ are called \textit{hyperedges}. A \textit{matching} of $\mathcal{H}$ is defined to be a collection of pairwise disjoint hyperedges of $\mathcal{H}$. The matching number, denoted by $\nu\left(\mathcal{H}\right)$, is the maximum size of a matching.

For our purpose, we define a hypergraph $\mathcal{H}_{q,n}=\left(\Sigma_q^{n-1},\mathcal{E}_{q,n}\right)$, where $\mathcal{E}_{q,n}=\left\{D_1^{(0)}(\bfx)~:~\bfx\in\Sigma_q^n\setminus\mathcal{B}_n\right\}$. Here $D_1^{(0)}(\bfx)\subseteq\Sigma_q^{n-1}$ is the set of sequences obtained by deleting exactly one zero from $\bfx$. For example, if $\bfx=0110010111$, then $D_1^{(0)}(\bfx)=\{110010111,011010111,011001111\}$. Obviously, a set $\mathcal{C}\subseteq\Sigma_q^n\setminus\mathcal{B}_n$ is a zero-deletion correcting code if and only if $\left\{D_1^{(0)}(\bfx)~:~\bfx\in\mathcal{C}\right\}$ is a matching of $\mathcal{H}_{q,n}$, and hence $A_{q,n}=\nu\left(\mathcal{H}_{q,n}\right)$. Therefore, the problem boils down to estimating $\nu\left(\mathcal{H}_{q,n}\right)$.

Suppose that $\mathcal{H}=\left(V,\mathcal{E}\right)$ is a hypergraph with $V=\{v_1,\ldots,v_n\}$ and $\mathcal{E}=\{E_1,\ldots,E_m\}$. Then the \textit{incidence} matrix $A$ of $\mathcal{H}$ is of size $n\times m$ and is defined as follows:
\begin{equation*}
  A_{i,j}=
  \begin{cases}
    1, & \mbox{if } v_i\in E_j, \\
    0, & \mbox{otherwise}.
  \end{cases}
\end{equation*}
Here $A_{i,j}$ is the element in the $i$th row and $j$th column of $A$.

The following lemma gives an upper bound of $\nu\left(\mathcal{H}\right)$.
\begin{lemma}\cite[Lemma 2.4]{Kulkarni2013}\label{lem_transversal}
  Let notations be as above. Then $\nu\left(\mathcal{H}\right)\le\tau^{*}\left(\mathcal{H}\right)$, where
  \begin{equation*}
    \tau^{*}\left(\mathcal{H}\right)=\min\left\{\mathop{\sum}_{i=1}^{n}w_i~:~A^T\bm{w}\ge \bm{1},\bm{w}\ge\bm{0}\right\}.
  \end{equation*}
Here $A^T$ denotes the transpose of the matrix $A$, $\bm{w}=(w_1,\ldots,w_n)^T$ is a column vector whose components are all nonnegative reals, $\bm{1}$ denotes the column vector whose components are all $1$, $\bm{0}$ denotes the column vector whose components are all $0$, and the inequalities are defined component-wise. 
\end{lemma}
According to Lemma \ref{lem_transversal}, we have $A_{q,n}=\nu\left(\mathcal{H}_{q,n}\right)\le\tau^{*}\left(\mathcal{H}_{q,n}\right)$. By definition,
\begin{equation*}
\begin{array}{rl}
\tau^{*}\left(\mathcal{H}_{q,n}\right)&=\min\left\{\mathop{\sum}\limits_{\bby\in\Sigma_q^{n-1}}w(\bby)~:~\mathop{\sum}_{\bby\in D_1^{(0)}(\bfx)}w(\bby)\ge 1,\forall\bfx\in\Sigma_q^n\setminus\mathcal{B}_n, \right.\\
&\quad\quad\quad\quad\quad \text{ and } \bm{w}(\bby)\ge\bm{0},\forall{\bby}\in\Sigma_q^{n-1}\Bigg\}.
\end{array}
\end{equation*}

For a sequence $\bbz\in\bigcup_{i=1}^{\infty}\Sigma_q^i$ of finite length, we let $r_0(\bbz)$ be the number of runs of zeros in $\bbz$. For example, if $\bbz=0110010111$, then $r_0(\bbz)=3$. It is clear that $r_0(\bby)\le r_0(\bfx)$ if $\bby\in D_1^{(0)}(\bfx)$.
If $\bby\in\mathcal{B}_{n-1}$, we let $w(\bby)=1$; otherwise, let $w(\bby)=\frac{1}{r_0(\bby)}$. Then $w(\bby)\ge 0$ and
\begin{equation*}
  \mathop{\sum}_{\bby\in D_1^{(0)}(\bfx)}w(\bby)\ge \mathop{\sum}_{\bby\in D_1^{(0)}(\bfx)}\frac{1}{r_0(\bfx)}=\frac{\left|D_1^{(0)}(\bfx)\right|}{r_0(\bfx)}=1
\end{equation*}
for any $\bfx\in\Sigma_q^n\setminus\mathcal{B}_n$. The last equality follows from the fact $\abs{D_1^{(0)}\parenv{\bfx}}=r_0\parenv{\bfx}$. Let $\mathcal{S}=\Sigma_q^{n-1}\setminus\mathcal{B}_{n-1}$. Since
\begin{equation}\label{eq_selectedvalue}
\mathop{\sum}\limits_{\bby\in\Sigma_q^{n-1}}w(\bby)=(q-1)^{n-1}+\mathop{\sum}\limits_{\bby\in \mathcal{S}}\frac{1}{r_0(\bby)}, \end{equation}
it remains to calculate $\sum_{\bby\in \mathcal{S}}\frac{1}{r_0(\bby)}$. Note that $1\le r_0(\bby)\le\left\lceil\frac{n-1}{2}\right\rceil=\left\lfloor\frac{n}{2}\right\rfloor$ for any $\bby\in \mathcal{S}$. However, these bounds are too loose and will only lead to $A_{q,n}\le q^{n-1}$. Thus, a better bound is needed. 

\begin{lemma}\label{lem_numofsolution}
	For a given positive integer $N$, the number of integer solutions to the following equation
	\begin{equation*}
		a_1+\cdots +a_t=N
	\end{equation*}
	under the condition that $a_i\ge 0$ for all $i=1,\ldots,t$, is $\binom{N+t-1}{N}$. More generally, the number of integer solutions to the above equation under the condition that $a_i\ge p_i$ for all $i=1,\ldots,t$, is $\binom{N+t-(\sum_{i=1}^{t}p_i)-1}{t-1}$, where $p_1,\ldots,p_t$ are nonnegative integers.
\end{lemma}
\begin{IEEEproof}
	The first conclusion is \cite[Proposition 1.5]{Jukna2011}. To prove the general conclusion, let $a_i^{\prime}=a_i-p_i$ for each $1\le i\le t$. Then each $a_i^{\prime}$ is a nonnegative integer. The proof follows from the first conclusion.
\end{IEEEproof}

\begin{lemma}\label{lem_numofy}
Let $n\ge 2$ be a positive integer. For any $1\le k\le\left\lfloor\frac{n}{2}\right\rfloor$, the number of sequences $\bby$ in $\mathcal{S}$ with the property $r_0(\bby)=k$ is $\binom{n-2}{2k}(q-1)^{k+1}+2\binom{n-2}{2k-1}(q-1)^{k}+\binom{n-2}{2k-2}(q-1)^{k-1}$.
\end{lemma}
\begin{IEEEproof}
For $\bby\in \mathcal{S}$ with $r_0(\bby)=k$, we can write $\bby$ in the form
$$
\bby=a_0^{m_0}0^{l_1}a_1^{m_1}\cdots 0^{l_k}a_k^{m_k},
$$
where $a_0,a_1,\ldots,a_k\in\Sigma_q\setminus\{0\}$, $m_0,m_k\ge 0$, $l_i,m_j\ge 1$ for all $1\le i\le k$ and $1\le j\le k-1$. Let $S_n$ be the number of solutions to the equation $\sum_{i=1}^{k}l_i+\sum_{j=0}^{k}m_j=n-1$. The following conclusions are clear from Lemma \ref{lem_numofsolution}:
\begin{itemize}
  \item if $m_0,m_k\ge 1$, $S_n=\binom{n-2}{2k}$;
  \item if $m_0=0,m_k=1$ or $m_0=1,m_k=0$, $S_n=\binom{n-2}{2k-1}$;
  \item if $m_0=m_k=0$, $S_n=\binom{n-2}{2k-2}$.
\end{itemize}
Therefore, the number of sequences is $\abs{\cS}=\binom{n-2}{2k}(q-1)^{k+1}+2\binom{n-2}{2k-1}(q-1)^{k}+\binom{n-2}{2k-2}(q-1)^{k-1}$.
\end{IEEEproof}

From Lemma \ref{lem_numofy} and \Cref{eq_selectedvalue}, we have
\begin{align}\label{eq_lastvalue}
\mathop{\sum}\limits_{\bm{y}\in\Sigma_q^{n-1}}w(\bm{y})&=(q-1)^{n-1}+\mathop{\sum}\limits_{\bm{y}\in \mathcal{S}}\frac{1}{r_0(\bm{y})}\notag\\
  &=(q-1)^{n-1}+\mathop{\sum}\limits_{k=1}^{\left\lfloor\frac{n}{2}\right\rfloor}\frac{1}{k}\binom{n-2}{2k}(q-1)^{k+1}\\
  &+2\mathop{\sum}\limits_{k=1}^{\left\lfloor\frac{n}{2}\right\rfloor}\frac{1}{k}\binom{n-2}{2k-1}(q-1)^{k}+\mathop{\sum}\limits_{k=1}^{\left\lfloor\frac{n}{2}\right\rfloor}\frac{1}{k}\binom{n-2}{2k-2}(q-1)^{k-1}\notag
\end{align}

To derive our desired result, we need the following lemma.
\begin{lemma}\cite[Claim 2]{Yaakobi2020isit}\label{lem_inequality}
  For integers $q\ge2$, $n\ge5$ and $n\ge q$, it holds that
  $$
  \sum_{k=1}^{n}\frac{1}{k}\binom{n}{k}(q-1)^k\le\frac{q^{n+1}}{(q-1)(n-2)}.
  $$
\end{lemma}

Putting everything together, we can now present the main theorem of this section.
\begin{theorem}
  Let notations be as above. For integers $q\ge 2$, $n\ge 12$ and $n\ge q$, it holds that $A_{q,n}\le (q-1)^{n-1}+1+\frac{8q^{n-1}}{(q-1)(n-4)}$. In particular, the redundancy of $\mathcal{C}_{max}$ is at least $\log_q(n)-\log_q(C_q)$, where $C_q$ is a constant dependent on $q$ and independent of $n$.
\end{theorem}
\begin{IEEEproof}
For any $k\ge 1$, we have $k+1\le 2k$. Therefore,
\begin{align}\label{eq_au1}
   \mathop{\sum}\limits_{k=1}^{\left\lfloor\frac{n}{2}\right\rfloor}\frac{1}{k}\binom{n-2}{2k}(q-1)^{k+1}\le 2\mathop{\sum}\limits_{k=1}^{\left\lfloor\frac{n}{2}\right\rfloor}\frac{1}{2k}\binom{n-2}{2k}(q-1)^{2k}\le 2\mathop{\sum}\limits_{k=1}^{n-2}\frac{1}{k}\binom{n-2}{k}(q-1)^{k}.
\end{align}
Since $1/k\le2/(2k-1)$ and $k\le 2k-1$, we have
\begin{align}\label{eq_au2}
    2\mathop{\sum}\limits_{k=1}^{\left\lfloor\frac{n}{2}\right\rfloor}\frac{1}{k}\binom{n-2}{2k-1}(q-1)^{k}\le 4\mathop{\sum}\limits_{k=1}^{\left\lfloor\frac{n}{2}\right\rfloor}\frac{1}{2k-1}\binom{n-2}{2k-1}(q-1)^{2k-1}\le 4\mathop{\sum}\limits_{k=1}^{n-2}\frac{1}{k}\binom{n-2}{k}(q-1)^{k}.
\end{align}
By \Cref{eq_au1}, we have
\begin{align}\label{eq_au3}
    \mathop{\sum}\limits_{k=1}^{\left\lfloor\frac{n}{2}\right\rfloor}\frac{1}{k}\binom{n-2}{2k-2}(q-1)^{k-1}&=1+\mathop{\sum}\limits_{k=2}^{\left\lfloor\frac{n}{2}\right\rfloor}\frac{1}{k}\binom{n-2}{2k-2}(q-1)^{k-1}\notag\\
    &\le 1+\mathop{\sum}\limits_{k=2}^{\left\lfloor\frac{n}{2}\right\rfloor}\frac{1}{k-1}\binom{n-2}{2k-2}(q-1)^{k-1}\notag\\
    &=1+\mathop{\sum}\limits_{k=1}^{\floorenv{\frac{n}{2}}-1}\frac{1}{k}\binom{n-2}{2k}(q-1)^{k}\notag\\
    &\le 1+2\mathop{\sum}\limits_{k=1}^{n-2}\frac{1}{k}\binom{n-2}{k}(q-1)^{k}.
\end{align}

Now combining \Cref{eq_lastvalue}, \Lref{lem_inequality} and \Cref{eq_au1,eq_au2,eq_au3}, we obtain
  \begin{align*}
  \mathop{\sum}\limits_{\bm{y}\in\Sigma_q^{n-1}}w(\bm{y})&=(q-1)^{n-1}+\mathop{\sum}\limits_{k=1}^{\left\lfloor\frac{n}{2}\right\rfloor}\frac{1}{k}\binom{n-2}{2k}(q-1)^{k+1}\\
  &+2\mathop{\sum}\limits_{k=1}^{\left\lfloor\frac{n}{2}\right\rfloor}\frac{1}{k}\binom{n-2}{2k-1}(q-1)^{k}+\mathop{\sum}\limits_{k=1}^{\left\lfloor\frac{n}{2}\right\rfloor}\frac{1}{k}\binom{n-2}{2k-2}(q-1)^{k-1}\\
  &\le (q-1)^{n-1}+1+2\mathop{\sum}\limits_{k=1}^{n-2}\frac{1}{k}\binom{n-2}{k}(q-1)^{k}\\
  &+4\mathop{\sum}\limits_{k=1}^{n-2}\frac{1}{k}\binom{n-2}{k}(q-1)^{k}+2\mathop{\sum}\limits_{k=1}^{n-2}\frac{1}{k}\binom{n-2}{k}(q-1)^{k}\\
  &\le (q-1)^{n-1}+1+\frac{8q^{n-1}}{(q-1)(n-4)}.
\end{align*}
By our discussion at the beginning of this section, we have $|\mathcal{C}_{max}|\le (q-1)^{n-1}+1+\frac{8q^{n-1}}{(q-1)(n-4)}+(q-1)^n$. When $n$ is large enough, this implies $|\mathcal{C}_{max}|\le C_q\frac{q^n}{n}$, where $C_q$ is a constant that depends on $q$ and independent of $n$. Therefore, the redundancy of $\mathcal{C}_{max}$ is at least $\log_q(n)-\log_q(C_q)$.
\end{IEEEproof}

\begin{corollary}
  The code in \Cref{eq_vtcode} (when $a=0$) is optimal up to a constant and the code in \Cref{thm_nonbinarycode} is optimal up to an $O(\log_q\log_q(n))$, in terms of redundancy.
\end{corollary}

\section{A Variant of the absorption channel and its connection with Deletion channels} 
\label{sec_variant}
In this section, we briefly discuss a variant of the absorption channel, which we call the \emph{contraction} channel. Interestingly, we find that it is equivalent to the deletion channel, which has been extensively studied in recent years.
Throughout this section, we assume that $q$ is a fixed positive integer great than $2$.
\begin{definition}\label{dfn_contraction}
  Suppose that $\bfx\in\Sigma_q^n$ is the transmitted sequence and $\bby\in\Sigma_q^{n-1}$ is the received sequence, where
\begin{itemize}
  \item $\bby=x_1\cdots x_{i-1}\parenv{x_i\boxplus x_{i+1}}x_{i+2}\cdots x_n$ for some $1\le i\le n-1$, or
  \item $\bby=x_1\cdots x_{n-1}$.
\end{itemize}
Here $x_i\boxplus x_{i+1}$ is defined to be $x_i+x_{i+1}\pmod{q}$.
For simplicity, in the rest of this section we will say that $\bby$ is obtained from $\bfx$ by a \emph{contraction} if $\bby$ is obtained from $\bfx$ in this way.
\end{definition}
With \Dref{dfn_contraction} in hand, multiple contractions can be defined in a similar way that we defined multiple absorptions (see \Dref{dfn_absorption}).

For any $\bfx\in\Sigma_q^n$ and any integer $t\in\sparenv{1,n-1}$, we define
$$
D_t\parenv{\bfx}=\mathset{\bbz\in\Sigma_q^{n-t}~:~\bbz\text{ is a subsequence of }\bfx}.
$$
Let $\cC$ be a nonempty subset of $\Sigma_q^n$. If $D_t\parenv{\bbc}\cap D_t\parenv{\bbc^{\prime}}=\emptyset$ for any two distinct sequences $\bbc,\bbc^{\prime}\in\cC$, we say $\cC$ is a $q$-ary $t$-deletion correcting code.
There are some known results on nonbinary $t$-deletion correcting codes with low redundancy \cite{Song2022IT,Sima2020isit_deletion}.

Next, We construct a bijection that connects between contractions and deletions. To that end, we use the following notation. For any $t\ge 0$ and $n\ge t+1$, let
$$
A_q\parenv{n,t}=\left\{\bfx\in\Sigma_q^n~:~x_i=0\text{ for all }1\le i\le t\right\}
$$
and
$$
B_q\parenv{n,t}=\left\{\bby\in\Sigma_q^{n+1}~:~y_i=0\text{ for all }1\le i\le t+1\right\}.
$$
We define a mapping $\Phi_{n,t}$ from $A_q\parenv{n,t}$ to $B_q\parenv{n,t}$ as following: 
\begin{equation*}
\begin{array}{rc}
   \Phi_{n,t}:& A_q\parenv{n,t}\rightarrow B_q\parenv{n,t}\\
   &\bfx\mapsto\bby
\end{array}
\end{equation*}
where $y_1=0$ and $y_i=\mathop{\boxplus}\limits_{j=1}^{i-1}x_j$ for each $i\ge 2$. Clearly, the mapping $\Phi_{n,t}$ is a bijection. Indeed, for any $\bby\in B_q\parenv{n,t}$, we have $\Phi_{n,t}^{-1}(\bby)=x_1\cdots x_n$, where $x_i=y_{i+1}-y_i\pmod{q}$ for each $1\le i\le n$.

\begin{lemma}\label{lem_equivalence}
  Let $\bfx\in A_q\parenv{n,t}$ and $\bby=\Phi_{n,t}(\bm{x})$, where $t$ is a positive integer and $n\ge t+1$ is an integer.
  \begin{enumerate}[$(1)$]
    \item $t$ contractions in $\bfx$ corresponds $t$ deletions in $\bby$.
    \item $t$ deletions in $\bby$ corresponds $t$ contractions in $\bfx$.
  \end{enumerate}
\end{lemma}

Before proving the lemma, we give a simple example to demonstrate the idea. 
\begin{example}
Let $n=7, t=1$ and consider the sequence $\bfx=0121201$ over the ternary alphabet $\Sigma_3=\mathset{0,1,2}$. Applying the bijection, we obtain $\Phi_{7,1}(\bfx)=\bby=00101001$. Now assume a contraction occurred in $\bfx$ in position $i=2$, i.e., we obtain $\bfx'=x_1 (x_2\boxplus x_3) x_4\dots x_7=001201$. The corresponding $\bby'=\Phi(\bfx')=0001001$ can be obtained from $\bby$ by deleting $y_3$. 

Considering $2$ consecutive contractions, let $\bfx''=01201$ be obtained by contracting $x_2\boxplus x_3 \boxplus x_4$. 
The corresponding $\bby'$ is $\Phi(\bfx'')=001001$ which can also be obtained by deleting $y_3$ and $y_4$ from $\bby$.
\end{example}

We now prove the lemma.
\begin{IEEEproof}
(1). Suppose that $\bfx^{\prime}$ is obtained from $\bfx$ by $t$ contractions. Then
\begin{equation}\label{eq_x}
x_i^{\prime}=
  \begin{cases}
    x_i, & \mbox{if }  i<i_{1},\\
    x_{i+\sum_{j=1}^{l}s_j}, & \begin{array}{l}\mbox{if } i_l-\sum_{j=1}^{l-1}s_j<i<i_{l+1}-\sum_{j=1}^{l}s_j\\
    \text{ for some }1\le l<k,\end{array}\\
    x_{i+\sum_{j=1}^{k}s_j}, & \mbox{if } i>i_k-\sum_{j=1}^{k-1}s_j,\\
    \mathop{\boxplus}\limits_{j=i_l}^{i_l+s_l}x_j, & \mbox{if } i=i_l-\sum_{j=1}^{l-1}s_j\text{ for some }1\le l\le k.
  \end{cases}
\end{equation}
Here $s_l\ge 1$ for each $1\le l\le k$, the sum $\sum_{l=1}^{k}s_l=t-t^{\prime}$, $i_1\ge 1$, $i_k+s_k\le n-t^{\prime}$ and $i_{l+1}-i_l>s_l$ for each $1\le l<k$. Let $\bby^{\prime}=\Phi_{n-t,0}\parenv{\bfx^{\prime}}$. Then $\bby^{\prime}$ is obtained from $\bby$ by deleting $y_{i_l+r}$ ($1\le l\le k,1\le r\le s_l$) and $\bby_{\sparenv{n-t^{\prime}+2,n+1}}$. Therefore, $\bby^{\prime}$ is obtained from $\bby$ by $t$ deletions.

(2). Suppose that $\bby^{\prime}$ is obtained from $\bby$ by $t$ deletions. Then there exist integers $i_l,s_l$ ($1\le l\le k$) satisfying $i_1\ge 0$, $s_l\ge 1$ for all $l\ge 1$, $i_{l+1}-i_l>s_l$ for all $1\le l<k$ and $i_k+s_k\le n-t^{\prime}$, such that $\bby^{\prime}$ is obtained from $\bby$ by deleting $y_{i_l+r}$ for all $1\le l\le k$ and $1\le r\le s_l$ (where $\sum_{l=1}^{k}s_l=t-t^{\prime}$) and $\bby_{\sparenv{n-t^{\prime}+2,n+1}}$. Notice that $y_1=\cdots=y_{t+1}=0$. So we can assume $i_1\ge 1$ and hence $y_1^{\prime}=0$. Let $\bfx^{\prime}=\Phi_{n-t,0}^{-1}\parenv{\bby^{\prime}}$. By construction, we can see that $\bfx^{\prime}$ is as in \Cref{eq_x}. Therefore, $\bfx^{\prime}$ is obtained from $\bfx$ by $t$ contractions.
\end{IEEEproof}

\Lref{lem_equivalence} suggests that a $t$-contraction error in sequences in $A_q(n,t)$ is equivalent to a $t$-deletion error in sequences in $B_q(n,t)$. Therefore, a $t$-contraction correcting code in $A_q(n,t)$ is equivalent to a $t$-deletion correcting code in $B_q(n,t)$.
\begin{observation}\label{obs_contraction}
Let $a,b\in\Sigma_q$, $a\ne b$, and $0<a,b\le q-1$.
\begin{itemize}
  \item If $x_ix_{i+1}\in\{0a,a0,00\}$, then $N_0(\bfx)=N_0(\bby)+1$ and $N_d(\bfx)=N_d(\bby)$ for all $d\ne 0$. In other words, $\bby$ is obtained from $\bfx$ by deleting one $0$.
  \item If $x_ix_{i+1}=aa$ and $c=a\boxplus a$, then $N_a(\bfx)=N_a(\bby)+2$, $N_c(\bfx)=N_c(\bby)-1$ and $N_d(\bfx)=N_d(\bby)$ for all $d\ne a,c$.
  \item If $x_ix_{i+1}=ab$ and $c=a\boxplus b$, then $N_a(\bfx)=N_a(\bby)+1$, $N_b(\bfx)=N_b(\bby)+1$, $N_c(\bfx)=N_c(\bby)-1$ and $N_d(\bfx)=N_d(\bby)$ for all $d\ne a,b,c$.
\end{itemize}
\end{observation}

With \Obsref{obs_contraction} in hand, it is easy to construct codes correcting contraction errors, as we did in \Cref{thm_primarycode}, \Cref{thm_nonbinarycode} and \Cref{thm_multipleerror}.  On the other hand, we can also construct codes via deletion correcting codes. Since these two kinds of constructions are straightforward, we omit the details.

\section{Conclusion}
\label{sec_conc}
In this paper, we introduced and studied absorption channels, which are closely related to neural communication systems. We constructed codes with near-optimal redundancy for single-absorption errors and codes with logarithmic redundancy for multiple-absorption errors. We also explored a variant of the absorption channels called contraction channels and showed that they are equivalent to deletion channels, which have numerous practical applications. We hope that this new finding will inspire new approaches to the construction of deletion-correcting codes.

In \Cref{sec_optimality}, we derived an upper bound on the size of single-absorption-correcting codes based on the fact that such codes must be able to correct the deletion of zeros. This bound implies that the redundancy of our single-absorption codes is optimal up to a constant or a term of $O(\log_q\log_q(n))$. However, this upper bound is not tight because a code that can correct a deletion of zeros is not necessarily a single-absorption-correcting code. Improving this upper bound would require a better estimate of the size of the $1$-absorption ball $\cB_1^{ab}\left(\bfx\right)$ (see \Cref{eq_absorptionball}) for each $\bfx$, which appears to be a difficult task because $\abs{\cB_1^{ab}\left(\bfx\right)}$ depends on the structure of $\bfx$. This problem is left for future research.
There are other interesting future research directions, which include 
\begin{itemize}
    \item deriving an upper bound on the size of multiple-absorption codes;
    \item finding new constructions of multiple-absorption codes;
    \item finding efficient encoders for $\cD\parenv{n;\bm{r},\bm{\alpha},\bm{\beta}}$ and $\cE\parenv{n;\bm{r},\bm{\alpha},\bm{\beta},a}$;
    \item exploring the general error model, in which a symbol's value may be decreased and the next symbol's value increased.
\end{itemize}

\appendices
\section{encoding and decoding algorithms for the set $\cR_{q,n+5}$}\label{sec_algorithms}
In this section, we will give an algorithm that encodes an arbitrary sequence $\bfx\in\Sigma_q^n$ into $\cR_{q,n+5}$. Since the encoding process is reversible, a decoding algorithm arises naturally.
Throughout this section, it is assumed that $(c_1-4)\log_q(e)/(4q^4)\ge 5$ and $c_2\log_q(e)/(4q^4)\ge 1$. For two finite sets $A$ and $B$, let $f:~A\rightarrow B$ be an injective mapping ($A$, $B$ and $f$ will be clear from the context). Then $f$ induces a bijection $f_A$ from $A$ to its image $f\parenv{A}$. By abuse of notations, we denote the inverse of $f_A$ by $f^{-1}$.

The basic idea of the encoding algorithm can be outlined as follows.
\begin{enumerate}
  \item Find two consecutive patterns $0011$ of distance larger than $\delta$.
  \item Delete a substring of length $\delta-4$ between these two patterns. This process aims to decrease the distance between these two patterns.
  \item Encode the position of this deleted substring and a compressed version of this substring into a block.
  \item Insert this block into another position to make sure that this insertion does not introduce two consecutive patterns of distance larger than $\delta$.
  \item Continue this process until there are no two consecutive patterns $0011$ of distance larger than $\delta$.
\end{enumerate}

First, we present a method to compress a length $\delta-4$ sequence that does not contain $0011$, into a shorter sequence. The following lemma follows similar ideas in \cite[Observation 1]{Bitar2021isit} and \cite[Proposition 1]{shuche2022arXiv}.

\begin{lemma}
Let $\cS$ be the set of all sequences of length $\delta-4$ that do not contain $0011$ as a substring. Then there exists an injective mapping $g:~\cS\rightarrow\Sigma_q^{\delta-\ceilenv{\log_q(n)}-9}$. Furthermore, the two mappings $g$ and $g^{-1}$ can be computed in $O\parenv{n}$ time.
\end{lemma}
\begin{IEEEproof}
Divide each $\bbs\in\cS$ into $(\delta-4)/4$ segments, each of length $4$. In other words, represent $\bbs$ as $\bbs=\bbs_1\bbs_2\cdots\bbs_{(\delta-4)/4}$, where $\bbs_i\in\Sigma_q^{4}$ for each $1\le i\le(\delta-4)/4$. Since $\bbs_i\ne0011$, there are at most $q^4-1$ choices of $\bbs_i$. This implies that each $\bbs_i$ can be represented by a symbol from the alphabet $\Sigma_{q^4-1}$, and a sequence $\bbs$ can be represented by a sequence $\bbu\in\Sigma_{q^4-1}^{(\delta-4)/4}$. Let $n_{\bbu}$ be the number of $q$-ary symbols to represent $\bbu$. Then
\begin{align*}
  n_{\bbu}&\le\ceilenv{\log_q\parenv{q^4-1}^{\frac{\delta-4}{4}}}\\
  &=\ceilenv{\delta-4+\frac{\delta-4}{4q^4}\log_q\parenv{1-\frac{1}{q^4}}^{q^4}}\\
  &\le\ceilenv{\delta-4-\frac{\delta-4}{4q^4}\log_q\parenv{e}}.
\end{align*}
The last inequality follows from the fact that the function $\parenv{1-1/x}^{x}$ is increasing in $x$ when $x>1$ and $\lim_{x\to\infty}(1-1/x)^x=1/e$. Since $(c_1-4)\log_q(e)/(4q^4)\ge 5$ and $c_2\log_q(e)/(4q^4)\ge 1$, we have $(\delta-4)\log_q\parenv{e}/(4q^4)\ge\ceilenv{\log_q(n)}+5$. So $n_{\bbu}\le\delta-\ceilenv{\log_q(n)}-9$. Recall that $c_1$ and $c_2$ are integers. Thus, the sequence $\bbu$ (and $\bbs$) can be represented by a $q$-ary sequence of length $\delta-\ceilenv{\log_q(n)}-9$.

The construction of $g$ (and $g^{-1}$) is straightforward. Since each $\bbs_i$ corresponds to a symbol from $\Sigma_{q^4-1}$, we can obtain $\bbu$ from $\bbs$ by replacing each $\bbs_i$ by the symbol from $\Sigma_{q^4-1}$ that corresponds to the value of its base-$q$ representation. We then transform $\bbu$ to a $q$-ary sequence $\bbv$ of length $\delta-\ceilenv{\log_q(n)}-9$. This can be done, for example, using a lookup table. Overall, transforming $\bbs\in\cS\subseteq \Sigma_q^{\delta-4}$ into $\bbv\in \Sigma_{q}^{\delta-\ceilenv{\log_q(n)}-9}$ can be done in $O\parenv{n}$ time. This process is reversible and $g^{-1}$ can be computed in $O\parenv{n}$ time.

\end{IEEEproof}

With this lemma, we describe our encoding algorithm in \Cref{alg_encoder}. 
We note that since $q^{\ceilenv{\log_q(n)}}\ge n$, there is an injective mapping from $\sparenv{2,n+1}$ to $\Sigma_q^{\ceilenv{\log_q(n)}}$. Let $b$ be such a mapping. By building a lookup table, the two mappings $b$ and $b^{-1}$ can be computed in $O(n)$ time. 

\Cref{alg_encoder} works as follows. We scan the sequence for $0011$ starting from the end of the sequence and going backward. If there is a block between two consecutive appearances of $0011$ which is longer than $\delta-4$, the length-$\parenv{\delta-4}$ suffix of that block is removed, compressed, and placed at the beginning of the sequences together with a pointer to its position and with $0011$ appended to it. 

\begin{algorithm}[!htbp]
\DontPrintSemicolon
\SetAlgoLined
\KwIn {$\bfx\in\Sigma_q^n$}
\KwOut {$\bbc=\Enc{\bfx}\in\cR_{q,n+5}$}
\textbf{Initialization}\;
$\bbc\gets 1\bfx0011$, $i\gets n+5$, $d\gets 1$\;
\While{$i\ge d+\delta$}
{
\eIf{there is no $j\in\sparenv{d+3,i-4}$ such that $\bbc_{\sparenv{j-3,j}}=0011$}{
$j\gets d-1$
}{
find the largest $j\in\sparenv{d+3,i-4}$ such that $\bbc_{\sparenv{j-3,j}}=0011$
}
\eIf{$i-j\le\delta$}{$i\gets j$}{
$\bbc\gets 0b\parenv{i-4}g\parenv{\bbc_{\sparenv{i-\delta+1,i-4}}}0011\bbc_{\sparenv{1,i-\delta}}\bbc_{\sparenv{i-3,n+5}}$\;
$d\gets d+\delta-4$
}
}
\Return{$\bbc$}
\caption{Encoding an arbitrary sequence of length $n$ into $\cR_{q,n+5}$}
\label{alg_encoder}
\end{algorithm}

\begin{algorithm}[!htbp]
\DontPrintSemicolon
\SetAlgoLined
\KwIn {$\Enc{\bfx}\in\cR_{q,n+5}$}
\KwOut {$\bfx$}
\textbf{Initialization}\;
$\hat{\bfx}\gets\Enc{\bfx}$\;
\While{$\hat{x}_1=0$}
{
$ind\gets b^{-1}\parenv{\hat{\bfx}_{\sparenv{2,\ceilenv{\log_q(n)}+1}}}$\;
$\hat{\bfx}\gets \hat{\bfx}_{\sparenv{\delta-3,ind}}g^{-1}\parenv{\hat{\bfx}_{\sparenv{\ceilenv{\log_q(n)}+2,\delta-8}}}\hat{\bfx}_{\sparenv{ind+1,n+5}}$
}
$\hat{\bfx}\gets\hat{\bfx}_{\sparenv{2,n+1}}$\;
\Return{$\hat{\bfx}$}
\caption{Decoding $\Enc{\bfx}\in\cR_{q,n+5}$ into $\bfx$}
\label{alg_decoder}
\end{algorithm}

\begin{theorem}
  Given any sequence $\bfx\in\Sigma_q^n$, \Cref{alg_encoder} outputs a sequence $\Enc{\bfx}\in\cR_{q,n+5}$.
\end{theorem}

\begin{IEEEproof}
We start with a detailed explanation of the idea behind \Cref{alg_encoder}. In the Initialization step, a pattern $0011$ is appended to the end of the input sequence $\bfx$ since each sequence in $\cR_{q,n+5}$ ends with $0011$, and $1$ is appended to the beginning of $\bfx$. This appended $1$ serves as a marker for the beginning of the information sequence (or, alternatively, when to finish the decoding process). The variable $d$ is a pointer to the position of this symbol. The index $i$ is initialized to be $n+5$, which is the position of the last pattern $0011$ in $\bbc$. The condition for continuing the while loop is $i\ge d+\delta$. This is because we want to find two consecutive patterns $0011$ of distance larger than $\delta$.

The idea for the while loop is to search patterns $0011$ in the sequence, starting from the end of the sequence and going backward. Once the pattern $0011$ is encountered at position $i$ (i.e., the position of the last symbol in $0011$ is $i$), we search for the next pattern $0011$ that is closest to the one at position $i$. Assume there is a $0011$ pattern in position $j<i$ (the position of the last symbol is $j$). If the distance between these two patterns is at most $\delta$, we set $j\to i$ and repeat the process. Otherwise, we delete the substring $\bbc_{\sparenv{i-\delta+1,i-4}}$ of length $\delta-4$ and then insert a block $0b\parenv{i-4}g\parenv{\bbc_{\sparenv{i-\delta+1,i-4}}}0011$ at the beginning. This block contains the position $b(i-4)$ of the deleted substring and the compressed version $g\parenv{\bbc_{\sparenv{i-\delta+1,i-4}}}$ of the deleted substring. Notice that the length of the block $0b\parenv{i-4}g\parenv{\bbc_{\sparenv{i-\delta+1,i-4}}}0011$ is $\delta-4$. So the deletion-insertion process does not change the length of the input sequence.

Recall that in the Initialization step, a symbol $1$ was inserted at the beginning of the sequence and the variable $d$ denotes the position of this symbol. Since the inserted block is on the left of $c_d$ and the deleted substring is on the right of $c_d$, the value of $d$ should increase by $\delta-4$ in step 13. In steps 9--14, either $i$ decreases to $j$ or $d$ increases by $\delta-4$. So the while loop will end after a finite number of cycles. In other words, the algorithm will terminate after finite steps. In each loop, if two consecutive patterns of distance larger than $\delta$ are encountered, then the distance between them will decrease since a length $\delta-4$ substring between them is deleted. The distance of two existing consecutive patterns does not increase after the insertion of a block $0b\parenv{i-4}g\parenv{\bbc_{\sparenv{i-\delta+1,i-4}}}0011$. Besides, the insertion of a block will not introduce two consecutive patterns of distance larger than $\delta$ since the length of each block is $\delta-4$ and each block ends with $0011$.
So in the output sequence $\Enc{\bfx}$, the distance between two consecutive patterns is at most $\delta$ and thus $\Enc{\bfx}\in\cR_{q,n+5}$.
\end{IEEEproof}

The time for searching $i$ and $j$ are both $O(n)$. The time for computing $g$ and $b$ are both $O(n)$. For each pair $(i,j)$, there are at most $O\parenv{n/\log_q(n)}$ substrings of length $\delta-4$ to be deleted. Therefore, the time complexity of \Cref{alg_encoder} is $O\parenv{n^4/\log_q(n)}$.

It is easy to see that the encoding process of \Cref{alg_encoder} is reversible. The decoding algorithm is presented in \Cref{alg_decoder}.
We give a brief explanation of the correctness of \Cref{alg_decoder}. In the Initialization step of \Cref{alg_encoder}, a symbol $1$ was inserted at the beginning. This $1$ was not destroyed during the encoding process. Each inserted block $0b\parenv{i-4}g\parenv{\bbc_{\sparenv{i-\delta+1,i-4}}}0011$ begins with $0$. So in \Cref{alg_decoder}, the condition $\hat{x}_1=0$ implies that $\hat{\bfx}$ should be decoded. If $\hat{x}_1=1$ (this is exactly the inserted $1$), we just need to delete the first and the last four symbols in $\hat{\bfx}$. The remaining substring $\hat{\bfx}_{\sparenv{2,n+1}}$ is the original sequence $\bfx$. The time complexity of \Cref{alg_decoder} is $O\parenv{n^2}$.

\section{Proof of \Tref{thm_window}}\label{appendix1}
If the pattern $0011$ in the end of $\bbz_{l_{\bfx}}^{\bfx}$ was destroyed, then this error is easy to detect and correct, since each codeword $\bfx\in\mathcal{D}_1$ ends with $0011$. Therefore, we always assume that the absorption error does not destroy the pattern $0011$ in the end of $\bbz_{l_{\bfx}}^{\bfx}$.

If $\abs{\bby}=n$, then $\bby$ is error-free. If $\abs{\bby}=n-1$, then a single absorption happened. Notice that by calculating $g(\bby)-r_2\pmod{3}$, we can find $l_{\bby}-l_{\bfx}$ (see \Obsref{obs_marker}).
\vspace{10pt}

\noindent\textbf{Case (1)}: $g(\bby)-r_2\equiv 0\pmod{3}$. In this case, we have $l_{\bby}=l_{\bfx}$ and so we can assume $\bbz^{\bby}=\parenv{\bbz_1^{\bfx},\ldots,\bbz_{i-1}^{\bfx},\bbz_i^{\prime},\bbz_{i+1}^{\bfx},\ldots,\bbz_{l_{\bfx}}^{\bfx}}$ for some $i\le l_{\bfx}$,
where $\bbz_i^{\prime}$ is obtained from $\bbz_i^{\bfx}$ by an absorption error and so $\abs{\bbz_i^{\prime}}=\abs{\bbz_i^{\bfx}}-1$. Therefore, we have
\begin{align*}
  f(\bfx)-f(\bby)\equiv\mathop{\sum}\limits_{j=1}^{l_{\bfx}}j\abs{\bbz_j^{\bfx}}-\mathop{\sum}\limits_{j=1}^{l_{\bby}}j\abs{\bbz_j^{\bby}}\equiv i\parenv{\abs{\bbz_i^{\bfx}}-\abs{\bbz_i^{\prime}}}\equiv i\pmod{2n}.
\end{align*}

Since $1\le i\le l_{\bfx}\le n/4$ and $i\equiv f(\bfx)-f(\bby)\equiv r_1-f(\bby)\pmod{2n}$, we can find the value of $i$ from $\parenv{r_1-f(\bby)}\pmod{2n}$. This gives a window $W$ of length at most $\delta=O(\log_q(n))$ in which the absorption error has occurred. Furthermore, since $\parenv{r_1-f(\bby)}\pmod{2n}$ can be computed in $O(n)$ time, this window can be found in $O(n)$ time.
\vspace{10pt}

\noindent\textbf{Case (2)}: $g(\bby)-r_2\equiv 2\pmod{3}$. In this case, we have $l_{\bby}=l_{\bfx}-1$ and so we can assume $\bbz^{\bby}=\parenv{\bbz_1^{\bfx},\ldots,\bbz_{i-1}^{\bfx},\bbz_i^{\prime},\bbz_{i+2}^{\bfx},\ldots,\bbz_{l_{\bfx}}^{\bfx}}$ for some $i<l_{\bfx}$,
where $\bbz_i^{\prime}$ is obtained from $\bbz_i^{\bfx}$ and $\bbz_{i+1}^{\bfx}$ by an absorption error which destroyed the $0011$ in $\bbz_i^{\bfx}$ and so $\abs{\bbz_i^{\prime}}=\abs{\bbz_i^{\bfx}}+\abs{\bbz_{i+1}^{\bfx}}-1$. Therefore, we have
\begin{align*}
  f(\bfx)-f(\bby)&=\mathop{\sum}\limits_{j=1}^{l_{\bfx}}j\abs{\bbz_j^{\bfx}}-\mathop{\sum}\limits_{j=1}^{l_{\bby}}j\abs{\bbz_j^{\bby}}\pmod{2n}\\
  &=i\abs{\bbz_i^{\bfx}}+(i+1)\abs{\bbz_{i+1}^{\bfx}}-i\abs{\bbz_i^{\prime}}+\mathop{\sum}\limits_{j=i+2}^{l_{\bfx}}\abs{\bbz_j^{\bfx}}\pmod{2n}\\
  &=i+\abs{\bbz_{i+1}^{\bfx}}+ \mathop{\sum}\limits_{j=i+2}^{l_{\bfx}}\abs{\bbz_j^{\bfx}}\pmod{2n}.
\end{align*}
Since $0<i+\abs{\bbz_{i+1}^{\bfx}}+ \mathop{\sum}\limits_{j=i+2}^{l_{\bfx}}\abs{\bbz_j^{\bfx}}<\mathop{\sum}\limits_{j=1}^{l_{\bfx}}\abs{\bbz_j^{\bfx}}=n$, we can obtain the value of $i+\abs{\bbz_{i+1}^{\bfx}}+ \mathop{\sum}\limits_{j=i+2}^{l_{\bfx}}\abs{\bbz_j^{\bfx}}$ from $\parenv{r_1-f(\bby)}\pmod{2n}$.

For each $i\le i^{\prime}\le l_{\bby}$, we define
\begin{align*}
\Phi(i^{\prime})=\mathop{\sum}\limits_{j=i^{\prime}+1}^{l_{\bby}}\abs{\bbz_j^{\bby}}+i^{\prime}=\mathop{\sum}\limits_{j=i^{\prime}+2}^{l_{\bfx}}\abs{\bbz_j^{\bfx}}+i^{\prime}.
\end{align*}
Then we have
\begin{align}\label{eq1}
  \abs{\Phi(i)-\parenv{f(\bfx)-f(\bby)}}=\abs{\bbz_{i+1}^{\bfx}}\le\delta.
\end{align}
Besides, since $\abs{\bbz_{j}^{\bfx}}\ge 4$ for all $j$, it holds that
\begin{equation}
\begin{aligned}
  \Phi(i^{\prime}-1)-\Phi(i^{\prime})=\abs{\bbz_{i^{\prime}+1}^{\bfx}}-1\ge 3, 
\end{aligned}
\end{equation}
whenever $i^{\prime}-1\ge i$,
which in turn, implies that for $k\in\N$ such that $i^{\prime}-k\geq i$, 
\begin{equation}\label{eq2}
\begin{aligned}
  \Phi(i^{\prime}-k)-\Phi(i^{\prime})=\sum_{j=i^{\prime}-k+2}^{i^{\prime}+1}\abs{\bbz_{j}^{\bfx}}-k\ge 3k. 
\end{aligned}
\end{equation}

Now we can recover the desired window $W$ in the following way. Sequentially compute $\Phi(i^{\prime})$ for $i^{\prime}$ starting at $l_{\bby}$ until we find an $i_0\ge i$ such that $\abs{\Phi(i_0)-\parenv{f(\bfx)-f(\bby)}}\le\delta$. This $i_0$ does exist due to \Cref{eq1}. We claim that $i_0-i\le \frac{2}{3}\delta$. Otherwise, \Cref{eq2} implies that
\begin{align*}
  \abs{\Phi(i)-\parenv{f(\bfx)-f(\bby)}}&=\abs{\Phi(i)-\Phi(i_0)+\Phi(i_0)-\parenv{f(\bfx)-f(\bby)}}\\
  &\ge\abs{\Phi(i_0)-\Phi(i)}-\abs{\Phi(i_0)-\parenv{f(\bfx)-f(\bby)}}\\
  &>3\frac{2}{3}\delta -\delta=\delta,
\end{align*}
which contradicts \Cref{eq1}. Since $\abs{\bbz_i^{\bby}}\le2\delta-1$, $\abs{\bbz_j^{\bby}}\le\delta$ for each $j\ne i$ and $i_0-i\le \frac{2}{3}\delta$, obtaining $i_0$ gives a window $W$ of length $\abs{W}\le\parenv{i_0-i+1}\delta+\delta-1\le \frac{2}{3}\delta^2+2\delta-1\le c_4\log_q^2(n)$ for some constant $c_4$ depending on $c_1$ and $c_2$. This window contains the position where the absorption error happened.
\vspace{10pt}

\noindent\textbf{Case (3)}: $g(\bby)-r_2\equiv 1\pmod{3}$. In this case, $l_{\bby}=l_{\bfx}+1$ and so we can assume $\bbz^{\bby}=\parenv{\bbz_1^{\bfx},\ldots,\bbz_{i-1}^{\bfx},\bbz_i^{\prime},\bbz_i^{\prime\prime},\bbz_{i+1}^{\bfx},\ldots,\bbz_{l_{\bfx}}^{\bfx}}$ for some $i\le l_{\bfx}$,
where $\bbz_i^{\prime}$ and $\bbz_i^{\prime\prime}$ are obtained from $\bbz_i^{\bfx}$ by an absorption error which created a new $0011$ in $\bbz_i^{\bfx}$ and so $\abs{\bbz_i^{\prime}}+\abs{\bbz_i^{\prime\prime}}=\abs{\bbz_i^{\bfx}}-1$. Therefore, we have
\begin{align*}
  f(\bfx)-f(\bby)&=\mathop{\sum}\limits_{j=1}^{l_{\bfx}}j\abs{\bbz_j^{\bfx}}-\mathop{\sum}\limits_{j=1}^{l_{\bby}}j\abs{\bbz_j^{\bby}}\pmod{2n}\\
  &=i\abs{\bbz_i^{\bfx}}-i\abs{\bbz_i^{\prime}}-(i+1)\abs{\bbz_{i}^{\prime\prime}}-\mathop{\sum}\limits_{j=i+1}^{l_{\bfx}}\abs{\bbz_j^{\bfx}}\pmod{2n}\\
  &=i-\abs{\bbz_{i}^{\prime\prime}}-\mathop{\sum}\limits_{j=i+1}^{l_{\bfx}}\abs{\bbz_j^{\bfx}}\pmod{2n}.
\end{align*}
Since $1\le i\le l_{\bfx}\le n/4$, $4\le\abs{\bbz_i^{\prime\prime}}\le\abs{\bbz_i^{\bfx}}-5$ and $4\le\abs{\bbz_j^{\bfx}}$, we have
$$
-(n-6)\le i-\abs{\bbz_{i}^{\prime\prime}}-\mathop{\sum}\limits_{j=i+1}^{l_{\bfx}}\abs{\bbz_j^{\bfx}}\le n/4-4.
$$
Here, $f(\bfx)-f(\bby)$ is chosen to be the unique integer $-n+6\le a\le n/4-4$ such that $f(\bfx)-f(\bby)\equiv a\pmod{2n}$. In fact, we have $a=i-\abs{\bbz_{i}^{\prime\prime}}-\mathop{\sum}\limits_{j=i+1}^{l_{\bfx}}\abs{\bbz_j^{\bfx}}$.

Similar to Case (2), for each $i\le i^{\prime}< l_{\bby}$, we define
\begin{align*}
  \Phi(i^{\prime})=-\mathop{\sum}\limits_{j=i^{\prime}+2}^{l_{\bby}}\abs{\bbz_j^{\bby}}+i^{\prime}=-\mathop{\sum}\limits_{j=i^{\prime}+1}^{l_{\bfx}}\abs{\bbz_j^{\bfx}}+i^{\prime}.
\end{align*}
Then we have
\begin{align}\label{eq3}
  \abs{\Phi(i)-\parenv{f(\bfx)-f(\bby)}}=\abs{\bbz_{i}^{\prime\prime}}=\abs{\bbz_i^{\bfx}}-1-\abs{\bbz_i^{\prime}}\le\delta-5.
\end{align}
Besides, since $\abs{\bbz_{j}^{\bfx}}\ge 4$ for all $j$, it holds that
\begin{equation}\label{eq4}
\begin{aligned}
  \Phi(i^{\prime})-\Phi(i^{\prime}-1)=\abs{\bbz_{i^{\prime}}^{\bfx}}+1\ge 5.
\end{aligned}
\end{equation}
whenever $i^{\prime}-1\ge i$,

Now we can recover the desired window $W$ in the following way. Sequentially compute $\Phi(i^{\prime})$ for $i^{\prime}$ starting at $l_{\bby}-1$ until we find an $i_0\ge i$ such that $\abs{\Phi(i_0)-\parenv{f(\bfx)-f(\bby)}}\le\delta-5$. This $i_0$ does exist due to \Cref{eq3}. We claim that $i_0-i\le \frac{2}{5}\delta$. Otherwise, \Cref{eq4} implies that
\begin{align*}
  \abs{\Phi(i)-\parenv{f(\bfx)-f(\bby)}}&=\abs{\Phi(i)-\Phi(i_0)+\Phi(i_0)-\parenv{f(\bfx)-f(\bby)}}\\
  &\ge\abs{\Phi(i_0)-\Phi(i)}-\abs{\Phi(i_0)-\parenv{f(\bfx)-f(\bby)}}\\
  &>2\delta-\delta=\delta,
\end{align*}
which contradicts \Cref{eq3}. Since $\abs{\bbz_j^{\bfx}}\le\delta$ for each $j$ and $i_0-i\le \frac{2}{5}\delta$, obtaining $i_0$ gives a window $W$ of length $\abs{W}\le\parenv{i_0-i+1}\delta\le \frac{2}{5}\delta^2+\delta\le c_5\log_q^2(n)$ for some constant $c_5$ depending on $c_1$ and $c_2$. This window contains the position where the absorption error happened.

In Case (2) and Case (3), $f(\bfx)-f(\bby)$ can be computed in linear time as the process for searching an $i_0$. Therefore, the window $W$ can be found in $O(n)$ time.
Now let $c_3=\max\mathset{c_4,c_5}$ and the proof is completed.

\section{Non-binary absorption-correcting codes with efficient encoders and decoders}\label{variantcodes}
In this section, by applying the results in \Cref{sec_improved} and \Cref{sec_multiple}, we give two new absorption-correcting codes that are polynomial-time encodable and decodable.

For a set $A$ of size $m$, there exists an injection $\cQ$ from $A$ to $\Sigma_q^{\ceilenv{\log_q(m)}}$. Under this mapping, each element $a$ in $A$ can be represented as a $q$-ary sequence $\cQ\parenv{a}$ of length $\ceilenv{\log_q(m)}$. By building a lookup table, $\cQ$ and $\cQ^{-1}$ can be cumputed in $O(m)$ time.

\subsection{Single-absorption correcting codes}
Let $f$, $g$, $\hat{g}_1$ and $\hat{g}_2$ be as in \Cref{sec_improved}. A message $\bfx\in\Sigma_q^n$ is encoded into 
$$
\cE_1\parenv{\bfx}=\Enc{\bfx}010f\parenv{\Enc{\bfx}}g\parenv{\Enc{\bfx}}\hat{g}_1\parenv{\Enc{\bfx}}\hat{g}_2\parenv{\Enc{\bfx}},
$$
where $\Enc{\cdot}$ is the encoder in \Cref{alg_encoder}. Here the sequence $f\parenv{\Enc{\bfx}}g\parenv{\Enc{\bfx}}\hat{g}_1\parenv{\Enc{\bfx}}\hat{g}_2\parenv{\Enc{\bfx}}$ is defined to be the sequence
$$
\cQ\parenv{\parenv{f\parenv{\Enc{\bfx}},g\parenv{\Enc{\bfx}},\hat{g}_1\parenv{\Enc{\bfx}},\hat{g}_2\parenv{\Enc{\bfx}}}}.
$$
Therefore, $f\parenv{\Enc{\bfx}}g\parenv{\Enc{\bfx}}\hat{g}_1\parenv{\Enc{\bfx}}\hat{g}_2\parenv{\Enc{\bfx}}$ is a sequence of length $\log_q(n)+12\log_q\log_q(n)+O(1)$.

\begin{proposition}\label{prop_nonbinary1}
    The code 
    $
    \mathset{\cE_1\parenv{\bfx}:\bfx\in\Sigma_q^n}
    $
    is a single-absorption correcting code with redundancy $\log_q(n)+12\log_q\log_q(n)$ $+O(1)$.
\end{proposition}
\begin{IEEEproof}
The redundancy is clear from construction.
Denote the length of the code by $N$.
    Suppose that $\bbc=\cE_1\parenv{\bfx}$ is the transmitted codeword and $\hat{\bbc}$ is obtained from $\bbc$ by a single-absorption. Recall that the length of $\Enc{\bfx}$ is $n+5$. So $c_{\sparenv{n+6,n+8}}=010$. A single-absorption can not affect $c_{\sparenv{1,n+7}}$ and $c_{\sparenv{n+8,N}}$ simultaneously. Therefore, the decoder can recover $\bfx$ by the following procedure.
    \begin{itemize}
        \item If $\hat{c}_{n+6}=0$, no error occurred in $c_{\sparenv{1,n+7}}$ and so $\Enc{\bfx}=\hat{c}_{\sparenv{1,n+5}}$. Then the message $\bfx$ can be decoded from $\Enc{\bfx}$ by applying \Cref{alg_decoder}.
        \item If $\hat{c}_{n+6}=1$, an absorption occurred in $c_{\sparenv{1,n+7}}$. If $\hat{c}_{n+5}\ne 0$, no error occurred in $c_{[1,n+5]}$ and so $\Enc{\bfx}=\hat{c}_{\sparenv{1,n+5}}$. If $\hat{c}_{n+5}=0$, then $\hat{c}_{n+5}$ is obtained from $\Enc{\bfx}$ by an absorption. Notice that no error occurred in $c_{\sparenv{n+8,N}}$. So we have $\hat{c}_{\sparenv{n+8,N-1}}=f\parenv{\Enc{\bfx}}g\parenv{\Enc{\bfx}}\hat{g}_1\parenv{\Enc{\bfx}}\hat{g}_2\parenv{\Enc{\bfx}}$. By \Cref{thm_nonbinarycode}, we can recover $\Enc{\bfx}$ from $\hat{c}_{\sparenv{1,n+4}}$ when given $f\parenv{\Enc{\bfx}}g\parenv{\Enc{\bfx}}\hat{g}_1\parenv{\Enc{\bfx}}\hat{g}_2\parenv{\Enc{\bfx}}$.  Again, the message $\bfx$ can be decoded from $\Enc{\bfx}$ by applying \Cref{alg_decoder}.
    \end{itemize} 
\end{IEEEproof}

Since $\Enc{\cdot}$ is a polynomial-time encoder and the four functions $f$, $g$, $\hat{g}_1$ and $\hat{g}_2$ can be computed in polynomial time, the code in Proposition \ref{prop_nonbinary1} provides a polynomial-time encoder. By \Cref{alg_decoder} and the proofs of Proposition \ref{prop_nonbinary1} and \Cref{thm_nonbinarycode}, we can see that this code can also be decoded in polynomial time.

\subsection{Multiple-absorption correcting codes}
The construction of multiple-absorption correcting codes is more complicated. We first need the following trivial observation. Recall that $\cB_t^{ab}(\bfx)$ denotes the $t$-absorption ball centered at x.
\begin{observation}\label{obs_appendix}
    Let $\bbc=\bbc_1\bbc_2$ be a sequence. We assume that
   $\bbc_1=\bbc_1^{\prime}0^t$ and $\bbc_2=0^t\bbc_2^{\prime}$, where $\bbc_1^{\prime}$ and $\bbc_2^{\prime}$ are substrings of length at least $t+1$. Suppose $\bbc_1=\bbc_{\sparenv{n_0+1,n_1}}$ and $\bbc_2=\bbc_{\sparenv{n_1+1,n_2}}$, where $0=n_0<n_1<n_2=\abs{\bbc}$.
   Then for any $\hat{\bbc}\in\cB_t^{ab}\parenv{\bbc}$, we have $\hat{\bbc}_{\sparenv{n_{i-1}+1,n_i-t}}\in\cB_t^{ab}\parenv{\bbc_i}$ for each $i=1,2$.
\end{observation}

Let $\cE_1(\cdot)$ be the encoder given in Proposition \ref{prop_nonbinary1}. Define
\begin{equation*}
    \cE=\mathset{\cE_1\parenv{\bfx}0^t~:~\bfx\in\Sigma_q^n}.
\end{equation*}
Then Proposition \ref{prop_nonbinary1} ensures that $\cE$ is a single-absorption correcting code. Denote the length of this code by $n_1$. Then $n_1=n+\log_q(n)+o\parenv{\log_q(n)}$.
Claim \ref{clm_confusionball} and the proof of Lemma \ref{lem_compression} (here $R(n_1)=R_{q,n_1}$ as defined in \Cref{eq_valueofRq}) imply that there is a mapping $\bar{f}$ from $\cE$ to $\Sigma_q^{(4t-2)\log_q(n)+o\parenv{\log_q(n)}}$, such that $\bar{f}(\bbu)\ne\bar{f}\parenv{\bbu^{\prime}}$ for any $\bbu\ne\bbu^{\prime}\in\cE$ and $\cB_t^{ab}(\bbu)\cap\cB_t^{ab}\parenv{\bbu^{\prime}}\ne\emptyset$. Furthermore, Remark \ref{rmk_barf} asserts that $\bar{f}$ can be computed in polynomial time.

Now we are ready to give our construction. In this construction, a message $\bfx\in\Sigma_q^n$ is encoded into
\begin{equation*}
    \cE_2\parenv{\bfx}=\cE_1\parenv{\bfx}0^t0^th\parenv{\bfx}\Red{q,m}{0^th(\bfx)}
\end{equation*}
where $h\parenv{\bfx}=\bar{f}\parenv{\cE_1\parenv{\bfx}0^t}$, $m$ is the length of $0^th(\bfx)$ and $\Red{q,m}{\cdot}$ is defined as in \Cref{eq_DScode}.

\begin{proposition}\label{prop_nonbinary2}
Let $t\ge2$ be fixed.
    The code 
    $
    \mathset{\cE_2\parenv{\bfx}:\bfx\in\Sigma_q^n}
    $
    is a $t$-absorption correcting code with redundancy $(4t-1)\log_q(n)+o\parenv{\log_q(n)}$.
\end{proposition}
\begin{IEEEproof}
    The redundancy is clear from construction. Denote the length of this code by $n_2$. Suppose that $\bbc=\cE_2\parenv{\bfx}$ is the transmitted codeword and $\hat{\bbc}$ is obtained from $\bbc$ by a $t$ absorptions. By Observation \ref{obs_appendix}, we have $\hat{\bbc}_{\sparenv{1,n_1-t}}\in\cB_t^{ab}\parenv{\cE_1(\bfx)0^t}$ and $\hat{\bbc}_{\sparenv{n_1+1,n_2-t}}\in\cB_t^{ab}\parenv{0^th(\bfx)\Red{q,m}{0^th(\bfx)}}$. According to Lemma \ref{lem_DScode1} and Lemma \ref{lem_DScode2}, we can first recover $h(\bfx)$ from $\hat{\bbc}_{\sparenv{n_1+1,n_2-t}}$. Then Claim \ref{clm_confusionball} (this claim holds for any single-absorption code) and the property of $\bar{f}$ ensures that we can recover $\cE_1(\bfx)$ and thus $\bfx$ in polynomial time by brute force searching..
\end{IEEEproof}

Recall that $\cE_1(\bfx)$ and $h(\bfx)$ can be computed in polynomial time.
From Lemma \ref{lem_DScode1} and Lemma \ref{lem_DScode2}, we know that $\Red{q,m}{0^th(\bfx)}$ can be computed in polynomial time. Therefore, the code in Proposition \ref{prop_nonbinary2} provides a polynomial-time encoder. From the proof of Proposition \ref{prop_nonbinary2}, we can see that this code can also be decoded in polynomial time.

\bibliographystyle{IEEEtran}
\bibliography{ref}

\begin{thebibliography}{10}
\providecommand{\url}[1]{#1}
\csname url@samestyle\endcsname
\providecommand{\newblock}{\relax}
\providecommand{\bibinfo}[2]{#2}
\providecommand{\BIBentrySTDinterwordspacing}{\spaceskip=0pt\relax}
\providecommand{\BIBentryALTinterwordstretchfactor}{4}
\providecommand{\BIBentryALTinterwordspacing}{\spaceskip=\fontdimen2\font plus
\BIBentryALTinterwordstretchfactor\fontdimen3\font minus
  \fontdimen4\font\relax}
\providecommand{\BIBforeignlanguage}[2]{{%
\expandafter\ifx\csname l@#1\endcsname\relax
\typeout{** WARNING: IEEEtran.bst: No hyphenation pattern has been}%
\typeout{** loaded for the language `#1'. Using the pattern for}%
\typeout{** the default language instead.}%
\else
\language=\csname l@#1\endcsname
\fi
#2}}
\providecommand{\BIBdecl}{\relax}
\BIBdecl

\bibitem{Gabrys2022IT}
\BIBentryALTinterwordspacing
R.~Gabrys, V.~Guruswami, J.~Ribeiro, and K.~Wu, ``Beyond {S}ingle-{D}eletion
  {C}orrecting {C}odes: {S}ubstitutions and {T}ranspositions,'' \emph{IEEE
  Trans. Inf. Theory}, vol. Early Access, Aug. 2022. [Online]. Available:
  \url{https://ieeexplore.ieee.org/document/9869870}
\BIBentrySTDinterwordspacing

\bibitem{vashist2005smart}
S.~K. Vashist, R.~Tewari, I.~Kaur, R.~P. Bajpai, and L.~M. Bharadwaj,
  ``Smart-drug delivery system employing molecular motors,'' in \emph{Proc.
  Int. Conf. Intell. Sens. Inf. Process. (ICISIP)}, Chennai, India, Jan. 2005,
  pp. 441--446.

\bibitem{Davis1997}
S.~Davis, ``Biomedical applications of nanotechnology--implications for drug
  targeting and gene therapy,'' \emph{Trends Biotechnol.}, vol.~15, no.~6, pp.
  217--224, Jun. 1997.

\bibitem{dubach2007fluorescent}
J.~M. Dubach, D.~I. Harjes, and H.~A. Clark, ``Fluorescent {I}on-{S}elective
  {N}anosensors for {I}ntracellular {A}nalysis with {I}mproved {L}ifetime and
  {S}ize,'' \emph{Nano Lett.}, vol.~7, no.~6, pp. 1827--1831, Jun. 2007.

\bibitem{li2003cholesterol}
J.~Li, T.~Peng, and Y.~Peng, ``A {C}holesterol {B}iosensor {B}ased on
  {E}ntrapment of {C}holesterol {O}xidase in a {S}ilicic {S}ol-{G}el {M}atrix
  at a {P}russian {B}lue {M}odified {E}lectrode,'' \emph{Electroanalysis},
  vol.~15, no.~12, pp. 1031--1037, Jul. 2003.

\bibitem{tallury2010nanobioimaging}
P.~Tallury, A.~Malhotra, L.~M. Byrne, and S.~Santra, ``Nanobioimaging and
  sensing of infectious diseases,'' \emph{Adv. Drug Del. Rev.}, vol.~62, no.
  4-5, pp. 424--437, Mar. 2010.

\bibitem{yang2020comprehensive}
K.~Yang, D.~Bi, Y.~Deng, R.~Zhang, M.~M.~U. Rahman, N.~A. Ali, M.~A. Imran,
  J.~M. Jornet, Q.~H. Abbasi, and A.~Alomainy, ``A comprehensive survey on
  hybrid communication in context of molecular communication and terahertz
  communication for body-centric nanonetworks,'' \emph{IEEE Trans. Mol. Biol.
  Multi-Scale Commun.}, vol.~6, no.~2, pp. 107--133, Nov. 2020.

\bibitem{Akyildiz2008}
I.~F. Akyildiz, F.~Brunetti, and C.~Blázquez, ``Nanonetworks: {A} new
  communication paradigm,'' \emph{Comput. Networks}, vol.~52, no.~12, pp.
  2260--2279, Aug. 2008.

\bibitem{farsad2016comprehensive}
N.~Farsad, H.~B. Yilmaz, A.~Eckford, C.-B. Chae, and W.~Guo, ``A
  {C}omprehensive {S}urvey of {R}ecent {A}dvancements in {M}olecular
  {C}ommunication,'' \emph{IEEE Commun. Surv. Tutorials}, vol.~18, no.~3, pp.
  1887--1919, Thirdquater 2016.

\bibitem{pan2022molecular}
W.~Pan, X.~Chen, X.~Yang, N.~Zhao, L.~Meng, and F.~H. Shah, ``A {M}olecular
  {C}ommunication {P}latform {B}ased on {B}ody {A}rea {N}anonetwork,''
  \emph{Nanomaterials}, vol.~12, no.~4, p. 722, Feb. 2022.

\bibitem{Chen2011}
M.~Chen, S.~Gonzalez, A.~Vasilakos, H.~Cao, and V.~C.~M. Leung, ``Body {A}rea
  {N}etworks: {A} {S}urvey,'' \emph{Mobile Networks and Applications}, vol.~16,
  no.~2, pp. 171--193, Apr. 2011.

\bibitem{malak2012molecular}
D.~Malak and O.~B. Akan, ``Molecular communication nanonetworks inside human
  body,'' \emph{Nano Commun. Networks}, vol.~3, no.~1, pp. 19--35, Mar. 2012.

\bibitem{gerstner2002spiking}
W.~Gerstner and W.~M. Kistler, \emph{Spiking Neuron Models: Single Neurons,
  Populations, Plasticity}.\hskip 1em plus 0.5em minus 0.4em\relax Cambridge
  university press, 2002.

\bibitem{malak2014communication}
D.~Malak and O.~B. Akan, ``Communication theoretical understanding of
  intra-body nervous nanonetworks,'' \emph{IEEE Commun. Mag.}, vol.~52, no.~4,
  pp. 129--135, Apr. 2014.

\bibitem{akan2016fundamentals}
O.~B. Akan, H.~Ramezani, T.~Khan, N.~A. Abbasi, and M.~Kuscu, ``Fundamentals of
  {M}olecular {I}nformation and {C}ommunication {S}cience,'' \emph{Proceedings
  of the IEEE}, vol. 105, no.~2, pp. 306--318, Feb. 2017.

\bibitem{abbasi2018controlled}
N.~A. Abbasi, D.~Lafci, and O.~B. Akan, ``Controlled {I}nformation {T}ransfer
  {T}hrough {A}n {I}n {V}ivo {N}ervous {S}ystem,'' \emph{Sci. Rep.}, vol.~8,
  no.~1, pp. 1--12, Feb. 2018.

\bibitem{Silv}
D.~U. Silverthorn, \emph{Human Physiology : An Integrated Approach},
  8th~ed.\hskip 1em plus 0.5em minus 0.4em\relax Pearson, 2019.

\bibitem{hec2019}
R.~Heckel, G.~Mikutis, and R.~N. Grass, ``A {C}haracterization of the {DNA}
  {D}ata {S}torage {C}hannel,'' \emph{Sci. Rep.}, vol.~9, no.~1, pp. 1--12,
  Jul. 2019.

\bibitem{Yaakobi2020isit}
I.~Smagloy, L.~Welter, A.~Wachter-Zeh, and E.~Yaakobi, ``Single-{D}eletion
  {S}ingle-{S}ubstitution {C}orrecting {C}odes,'' in \emph{Proc. Int. Symp.
  Inf. Theory (ISIT)}, Los Angeles, CA, USA, Jun. 2020, pp. 775--780.

\bibitem{Song2022IT}
W.~Song, N.~Polyanskii, K.~Cai, and X.~He, ``Systematic {C}odes {C}orrecting
  {M}ultiple-{D}eletion and {M}ultiple-{S}ubstitution {E}rrors,'' \emph{IEEE
  Trans. Inf. Theory}, vol.~68, no.~10, pp. 6402--6416, Oct. 2022.

\bibitem{VL1966}
V.~I. Levenshtein, ``Binary codes capable of correcting deletions, insertions
  and reversals,'' \emph{Soviet Physics Doklady}, vol.~10, no.~8, pp. 707--710,
  Feb. 1966.

\bibitem{Sima2021it}
J.~Sima and J.~Bruck, ``On {O}ptimal $k$-{D}eletion {C}orrecting {C}odes,''
  \emph{IEEE Trans. Inf. Theory}, vol.~67, no.~6, pp. 3360--3375, Jun. 2021.

\bibitem{Guruswami2021it}
V.~Guruswami and J.~H{\aa}stad, ``Explicit {T}wo-{D}eletion {C}odes {W}ith
  {R}edundancy {M}atching the {E}xistential {B}ound,'' \emph{IEEE Trans. Inf.
  Theory}, vol.~67, no.~10, pp. 6384--6394, Oct. 2021.

\bibitem{Cheng2018FOCS}
K.~Cheng, Z.~Jin, X.~Li, and K.~Wu, ``Deterministic {D}ocument {E}xchange
  {P}rotocols, and {A}lmost {O}ptimal {B}inary {C}odes for {E}dit {E}rrors,''
  in \emph{Proc. Annu. Symp. Found. Comput. Sci. (FOCS)}, Paris, France, Oct.
  2018, pp. 200--211.

\bibitem{Haeupler2019FOCS}
B.~Haeupler, ``Optimal {D}ocument {E}xchange and {N}ew {C}odes for {I}nsertions
  and {D}eletions,'' in \emph{Proc. Annu. Symp. Found. Comput. Sci. (FOCS)},
  Baltimore, MD, USA, Nov. 2019, pp. 334--3--47.

\bibitem{VT1965}
R.~R. Varshamov and G.~M. Tenengolts, ``Code {C}orrecting {S}ingle {A}symmetric
  {E}rrors (in {R}ussian),'' \emph{Avtomat. i Telemekh.}, vol.~26, no.~2, pp.
  288--292, 1965.

\bibitem{Sloane2002}
N.~J.~A. Sloane, ``On single-deletion-correcting codes,'' \emph{Codes and
  Designs}, vol.~10, pp. 273--291, May 2002.

\bibitem{GhaFer1998}
K.~Abdel-Ghaffar and H.~Ferreira, ``Systematic encoding of the
  {V}arshamov-{T}enengol'ts codes and the {C}onstantin-{R}ao codes,''
  \emph{IEEE Trans. Inf. Theory}, vol.~44, no.~1, pp. 340--345, Jan. 1998.

\bibitem{Joshua2018it}
J.~Brakensiek, V.~Guruswami, and S.~Zbarsky, ``Efficient {L}ow-{R}edundancy
  {C}odes for {C}orrecting {M}ultiple {D}eletions,'' \emph{IEEE Trans. Inf.
  Theory}, vol.~64, no.~5, pp. 3403--3410, May 2018.

\bibitem{Gabrys2019it}
R.~Gabrys and F.~Sala, ``Codes {C}orrecting {T}wo {D}eletions,'' \emph{IEEE
  Trans. Inf. Theory}, vol.~65, no.~2, pp. 965--974, Feb. 2019.

\bibitem{Sima2020it}
J.~Sima, N.~Raviv, and J.~Bruck, ``Two {D}eletion {C}orrecting {C}odes {F}rom
  {I}ndicator {V}ectors,'' \emph{IEEE Trans. Inf. Theory}, vol.~66, no.~4, pp.
  2375--2391, Apr. 2020.

\bibitem{Tenengolts1984}
G.~Tenengolts, ``Nonbinary codes, correcting single deletion or insertion
  (corresp.),'' \emph{IEEE Trans. Inf. Theory}, vol.~30, no.~5, pp. 766--769,
  Sept. 1984.

\bibitem{Bruck2020isit}
J.~Sima, R.~Gabrys, and J.~Bruck, ``Optimal systematic $t$-deletion correcting
  codes,'' in \emph{Proc. Int. Symp. Inf. Theory (ISIT)}, Los Angeles, CA, USA,
  Jun. 2020, pp. 769--774.

\bibitem{Sima2019isit}
J.~Sima and J.~Bruck, ``Optimal $k$-{D}eletion {C}orrecting {C}odes,'' in
  \emph{Proc. Int. Symp. Inf. Theory (ISIT)}, Paris, France, Jul. 2019, pp.
  847--851.

\bibitem{Sima2020isit}
J.~Sima, R.~Gabrys, and J.~Bruck, ``Syndrome {C}ompression for {O}ptimal
  {R}edundancy {C}odes,'' in \emph{Proc. Int. Symp. Inf. Theory (ISIT)}, Los
  Angeles, CA, USA, Jun. 2020, pp. 751--756.

\bibitem{Kulkarni2013}
A.~A. Kulkarni and N.~Kiyavash, ``Nonasymptotic {U}pper {B}ounds for {D}eletion
  {C}orrecting {C}odes,'' \emph{IEEE Trans. Inf. Theory}, vol.~59, no.~8, pp.
  5115--5130, Aug. 2013.

\bibitem{Fazeli2015it}
A.~Fazeli, A.~Vardy, and E.~Yaakobi, ``Generalized {S}phere {P}acking
  {B}ound,'' \emph{IEEE Trans. Inf. Theory}, vol.~61, no.~5, pp. 2313--2334,
  Mar. 2015.

\bibitem{Cullina2016it}
D.~Cullina and N.~Kiyavash, ``Generalized sphere-packing bounds on the size of
  codes for combinatorial channels,'' \emph{IEEE Trans. Inf. Theory}, vol.~62,
  no.~8, pp. 4454--4465, May 2016.

\bibitem{Jukna2011}
S.~Jukna, \emph{Extremal Combinatorics}, 2nd~ed., ser. Texts in Theoretical
  Computer Science. An EATCS Series.\hskip 1em plus 0.5em minus 0.4em\relax
  Springer Berlin, Heidelberg, 2011.

\bibitem{Sima2020isit_deletion}
J.~Sima, R.~Gabrys, and J.~Bruck, ``Optimal {C}odes for the $q$-ary {D}eletion
  {C}hannel,'' in \emph{Proc. Int. Symp. Inf. Theory (ISIT)}, Los Angeles, CA,
  USA, Jun. 2020, pp. 740--745.

\end{thebibliography}

\end{document}